\documentclass[twocolumn]{aastex63}
\usepackage{amsmath}
\usepackage{amssymb}
\usepackage{multirow}

\newcommand{\bz}{$\langle B_z \rangle$}
\newcommand{\nz}{$\langle N_z \rangle$}

\newcommand{\teff}{$T_{\rm eff}$}

\usepackage{lineno}

\received{----}
\revised{----}
\accepted{----}
\submitjournal{ApJ}

\shortauthors{Das et al.}

\begin{document}

\title{Discovery of eight `Main-sequence Radio Pulse emitters' using the GMRT: clues to the onset of coherent radio emission in hot magnetic stars}

\correspondingauthor{Barnali Das}
\email{barnali@ncra.tifr.res.in}


\author[0000-0001-8704-1822]{Barnali Das}
\affil{National Centre for Radio Astrophysics, Tata Institute of Fundamental Research,  Pune University Campus, Pune-411007, India}

\author[0000-0002-0844-6563]{Poonam Chandra}
\affil{National Centre for Radio Astrophysics, Tata Institute of Fundamental Research,  Pune University Campus, Pune-411007, India}

\author{Matt E. Shultz}
\affiliation{Annie Jump Cannon fellow, Department of Physics and Astronomy, University of Delaware, 217 Sharp Lab, Newark, DE 19716, USA}

\author{Gregg A. Wade}
\affiliation{Department of Physics, Royal Military College of Canada, PO Box 17000, Station Forces, Kingston, ON K7K 7B4, Canada}

\author{James Sikora}
\affiliation{Bishop's University, Quebec J1M 1Z7, Canada}

\author{Oleg Kochukhov}
\affiliation{Department of Physics and Astronomy, Uppsala University, Box 516, SE-751 20 Uppsala, Sweden}

\author{Coralie Neiner}
\affiliation{LESIA, Paris Observatory, PSL University, CNRS, Sorbonne University, Univ.  Paris Diderot, Sorbonne Paris Cit, 5 place Jules Janssen, 92195 Meudon, France}

\author{Mary E. Oksala}
\affiliation{LESIA, Paris Observatory, PSL University, CNRS, Sorbonne University, Univ.  Paris Diderot, Sorbonne Paris Cit, 5 place Jules Janssen, 92195 Meudon, France}
\affiliation{Department of Physics, California Lutheran University,  60 W Olsen Rd, Thousand Oaks, CA 91360, USA}

\author{Evelyne Alecian}
\affiliation{ University  Grenoble Alpes, CNRS, IPAG, 38000 Grenoble, France}

\begin{abstract}
`Main-sequence radio pulse-emitters' (MRPs) are magnetic early-type stars from which periodic radio pulses, produced via electron cyclotron maser emission (ECME), are observed. Despite the fact that these stars can naturally offer suitable conditions for triggering ECME, only seven such stars have been reported so far within a span of more than two decades. In this paper, we report the discovery of eight more MRPs, thus more than doubling the sample size of such objects. These discoveries are the result of our sub-GHz observation program using the Giant Metrewave Radio Telescope over the years 2015--2021. Adding these stars to the previously known MRPs, we infer that at least 32\% of the magnetic hot stars exhibit this phenomenon, thus suggesting that observation of ECME is not a rare phenomenon. The significantly larger sample of MRPs allows us for the first time to perform a statistical analysis comparing their physical properties. We present an empirical relation that can be used to predict whether a magnetic hot star is likely to produce ECME. Our preliminary analysis suggests that the physical parameters that play the primary role in the efficiency of the phenomenon are the maximum surface magnetic field strength and the surface temperature. In addition, we present strong evidence of the influence of the plasma density distribution on ECME pulse profiles. 
Results of this kind further motivate the search for MRPs as a robust characterization of the relation between observed ECME properties and stellar physical parameters can only be achieved with a large sample.

\end{abstract}

\keywords{stars: magnetic field --- polarization --- masers}

\section{Introduction} \label{sec:intro}
Auroral radio emission (ARE) via electron cyclotron maser emission (ECME) has been observed from a wide variety of objects from early-type stars \citep[spectral type B or A, e.g.,][]{trigilio2000} to cool brown dwarfs and planets \citep[e.g.][]{hallinan2006}. The observation of this emission from the latter objects is highly useful as it is one of the best probes with which to estimate magnetic field strengths \citep[since the emission frequency is proportional to the local electron gyrofrequency, e.g.][]{melrose1982}. At the same time, such observations (from cool objects like ultracool dwarfs) are also curious as the emission requires not only a magnetic field, but also energetic electrons, the source of which is not always apparent. On the other hand, in the case of magnetic early-type stars, the situation is somewhat the opposite in the sense that they seem to have all the ingredients required for the production of ARE. Their magnetic fields often have $\sim$ kG strengths, have simple topologies \citep[usually well-approximated by dipoles, e.g.][]{kochukhov2019} and are also highly stable, exhibiting no sign of intrinsic change over at least thousands of rotational cycles \citep[e.g.][]{shultz2018}, and only gradually weakening over evolutionary timescales \citep[e.g.][]{landstreet2007,landstreet2008,sikora2019b,shultz2019b}.
The electrons required for ECM emission are supplied by the stellar wind (those that are energized within the magnetosphere). Moreover, the global dipole-like magnetic field can produce magnetic-mirror like conditions in which the electrons may undergo a `population inversion', necessary for maser emission \citep{trigilio2000,trigilio2004}. Nevertheless, only seven magnetic early-type stars have been observed to produce ECME: CU\,Vir \citep{trigilio2000}, HD\,133880 \citep{chandra2015,das2018}, HD\,142990 \citep{lenc2018,das2019a}, HD\,142301 \citep{leto2019}, HD\,35298 \citep{das2019b}, $\rho\,\mathrm{Oph A}$ \citep{leto2020a} and $\rho\,\mathrm{Oph C}$ \citep{leto2020b}. These ECME-producing magnetic early-type stars will be referred as `Main-sequence Radio Pulse emitters' \citep[MRPs,][]{das2021} as the emission is observed as period radio pulses.

In the past, there has been a suggestion that large deviations of the magnetic field from a dipolar geometry suppresses ECME \citep{leto2012}. This explanation is however inadequate since there are a number of MRPs, observed to produce ECME, for which the surface magnetic fields have been mapped via `Zeeman-Doppler Imaging' (ZDI) and shown to deviate significantly from that of an axi-symmetric dipole \citep[e.g. CU\,Vir and HD\,133880,][]{kochukhov2014,kochukhov2017}. 
Most recently, \citet{das2021} suggested that a complex magnetic field topology might affect the ECME beaming patterns so that even if a star produces ECME, and exhibits `magnetic null phases' (see \S\ref{sec:MRP_signature}), the radiation may not be visible to an observer over certain frequency ranges. The role of magnetic field topology on suppressing ECME is thus vague, and it therefore remains an open question why some stars become MRPs and others apparently do not.
One of the biggest hurdles in answering these questions is the very fact that only a small number of such stars are known.

In this paper, we report the discovery of eight new MRPs, more than doubling the size of the MRP population. Our discoveries result from sub-GHz observations carried out with the Giant Metrewave Radio telescope (GMRT) over the years 2015--2021. Combining these with already known MRPs, we, for the first time, present an empirical relation to predict whether a hot magnetic star is likely to produce ECME.

This paper is structured as follows: in the next section (\S\ref{sec:MRP_signature}), we explain what we expect to observe in our radio observations in order to identify an MRP candidate. This is followed by a brief description of the radio data acquisition and analysis (\S\ref{sec:radio_data}), and the results for individual stars (\S\ref{sec:results}). We discuss the results and summarize our conclusions in \S\ref{sec:discussion} and \S\ref{sec:conclusion} respectively.

\section{Signature of an `MRP'}\label{sec:MRP_signature}
In this section, we explain what we expect to see if the star is indeed an MRP. 
Radio emission from magnetic AB stars is primarily due to the gyrosynchrotron mechanism \citep[e.g.][]{drake1987}. Such emission (both total intensity and the percentage of circular polarization) smoothly varies with rotational phase, and the modulation correlates with that of the longitudinal magnetic field averaged over the visible stellar surface \bz~\citep[e.g.][]{leone1993,lim1996,leto2020a}. It has however been observed that the amplitude of the modulation decreases toward lower radio frequencies \citep[e.g.][]{leto2012,leto2020a}. As will be explained below, when looking for an MRP candidate, the primary signature is sharp variation of the flux density with rotational phase over a timescale much shorter than that characteristic of the basal gyrosynchrotron emission.

For a star with an oblique axi-symmetric dipolar magnetic field, ECME is produced in ring-shaped regions above the magnetic poles, called `auroral rings' \citep{trigilio2011}. The direction of emission is tangential to these rings such that the wave vector is perpendicular to the dipole axis \citep{trigilio2011}. As a result, the emission is expected to be seen when \bz~is zero (a magnetic null phase), corresponding to the magnetic equator bisecting the visible stellar disk. However, due to propagation effects in the magnetosphere, radiation from opposite magnetic hemispheres, which have opposite circular polarizations, get refracted. Hence, instead of a single pulse composed of radiation from both magnetic hemisphere, visible at the magnetic null phase, we expect to see a pair of oppositely circularly polarized pulses around each magnetic null phase \citep[e.g.][]{leto2016}.
The sequence of arrival of right and left circularly polarized \footnote{We follow the IAU/IEEE convention for circular polarization.} (RCP and LCP respectively) pulses are opposite near the two magnetic nulls \citep[see Figure 1 of][]{das2019a}. However, it has now become well known that such idealized behaviour is extremely rare. The rotational phases of arrival of the pulses often exhibit significant offsets from the nearest magnetic null phases \citep[][etc.]{trigilio2000, leto2020a}. Sometimes, pulses of only one circular polarization are observable \citep[e.g. CU\,Vir,][]{trigilio2000}. In the past, a very high brightness temperature ($T_\mathrm{B}\gtrsim 10^{12}\,\mathrm{K}$) and 100\% circular polarization were the two criteria for identification of coherent emission \citep{trigilio2000,das2018}. However, as shown by \citet{leto2016} and \citet{das2020a}, the observed circular polarization depends on propagation effects and under certain circumstances, can be zero as well. Thus very high circular polarization is not a necessary condition for the emission mechanism to be identified as ECME. In case of $T_\mathrm{B}$, one can only estimate a lower limit since the size of the emission region is not well-constrained \citep{das2018}. Setting the size of the emission site equal to the stellar disk, the expression for the brightness temperature is:

\begin{align}
    T_\mathrm{B}&\approx2\times 10^{13}\times \frac{Sd^2}{\nu^2R_*^2} \label{eq:T_B}
\end{align}
Where $S$ is the flux density (in mJy) observed at a frequency $\nu$ (in MHz) from a star with radius $R_*$ (in units of solar radii) at a distance of $d$ (in parsecs). Note that the actual source size is expected to be much smaller than the size of the stellar disk \citep[e.g.][]{trigilio2011}.
Therefore, even if the lower limit turns out to be within the limit of incoherent emission, one cannot use it to rule out ECME. On the other hand, if the lower limit to $T_\mathrm{B}$ is larger than $10^{12}$ K (the maximum allowed value for incoherent emission), it confirms the emission mechanism to be coherent. In such a case, ECME is almost always favoured over plasma emission as the latter cannot explain directed emission. Besides, to give rise to plasma emission at 0.6--0.8 GHz, the required number density is $\sim 10^9\,\mathrm{cm^{-3}}$, supposed to be present only at the densest part of the stellar magnetosphere \citep[close to the stellar surface,][]{leto2006,leto2020a}.

Based on the above facts, the only condition that we impose to identify an MRP candidate is observation of significant flux density enhancement over a rotational phase window that encompasses/is close to a magnetic null. The physical reason behind imposing this condition is that ECME is a highly directed phenomenon \citep[e.g.][]{melrose1982}. Since our observations were conducted at sub-GHz frequencies, where we do not expect to see much modulation due to gyrosynchrotron emission \citep[e.g.][]{leto2020a}, this condition is justified to identify a candidate.

In the ideal case where the gyrosynchrotron modulation follows that of \bz~(which, in the case of a dipole, varies sinusoidally with rotational phase $\phi_\mathrm{rot}$), we can define a necessary condition to attribute an enhancement to ECME. In this case, we approximate the variation of the gyrosynchrotron flux density $S$ as: $S(\phi_\mathrm{rot})=a\sin^2(2\pi\phi_\mathrm{rot})+b$, so that $b=S_\mathrm{min}$ and $a=S_\mathrm{max}-S_\mathrm{min}$, where $S_\mathrm{min}$ and $S_\mathrm{max}$ are the minimum and maximum values of $S(\phi_\mathrm{rot})$, respectively. The maximum gradient of the lightcurve then occurs at $\phi_\mathrm{rot}=0.125$, and the maximum value of this gradient is:
\begin{equation}
    \frac{dS}{d\phi_\mathrm{rot}}\biggl|_\mathrm{max}=2\pi a \label{eq:gyro_lightcurve_gradient}
\end{equation}
    
For the lightcurves presented in \S\ref{sec:results}, we calculate the quantity $\Delta S/\Delta\phi_\mathrm{rot}$, where $\Delta S$ is the change in flux density from the `base' to the peak of an enhancement over the rotational phase range $\Delta\phi_\mathrm{rot}$. In that case $\Delta S\sim a$, so that the necessary condition to attribute an enhancement to ECME becomes $1/\Delta\phi_\mathrm{rot}>2\pi$, or $\Delta\phi_\mathrm{rot}<0.159$. We refer to this condition as the `minimum flux density gradient condition'.

\section{Observations and data analysis}\label{sec:radio_data}
The eight stars in our sample, and their properties that are relevant for this study, are listed in Table \ref{tab:targets_properties}. 
In the next two subsections, we describe our selection criteria and observation strategy.

\subsection{Selection criteria}\label{subsec:selection_criteria}
All of the stars in our sample have well-characterized magneto-rotational properties. In addition, there are a few criteria that were applied to make our sample suitable for our science goal. These are listed below:

\begin{enumerate}
    \item Based on existing models, ECME is expected to be observable near magnetic null phases \citep[e.g.][]{trigilio2011,leto2016}. Hence, the first criterion that we imposed was that the \bz~modulation with rotational phase must have at least one magnetic null. Such a condition is realized for $i+\beta \geq 90^\circ$ (the equality condition corresponds to a single magnetic null), where $i$ is the angle between the line-of-sight and the rotation axis (inclination angle), and $\beta$ is the angle between the dipole and rotation axes (obliquity). 
   
    \item The next condition arises due to our choice of observing frequency. Our central frequency of observation (0.7 GHz) corresponds to a field strength of 250 G for ECME at the fundamental harmonic (using $\nu_\mathrm{B}\approx2.8 B$, where $\nu_\mathrm{B}$ is the electron gyrofrequency in MHz and $B$ is the local magnetic field strength in gauss). Until now, ECME upper cut-off frequencies have been reported for only two MRPs: HD\,133880 \citep{das2020b} and CU\,Vir \citep{das2021}. In both stars, the upper cut-off frequencies are significantly smaller than the electron gyrofrequency corresponding to the maximum magnetic field strength. For HD\,133880, the lowest height corresponding to the upper cut-off frequency is $\approx 0.6\,R_*$ \citep{das2020b}. In case of CU\,Vir, it was not possible to estimate the height due to the complex lightcurves \citep{das2021}. Assuming $0.6\,R_*$ as the minimum height for ECME production, we find that the surface polar strength of the star should be $>1\,\mathrm{kG}$ (for emission at the fundamental harmonic). All of the stars in our sample satisfy this condition.
    
    \item For observational convenience, we limited the survey to rapid rotators (rotation periods $P_\mathrm{rot}<2$ days). The only exception is HD\,79158 with $P_\mathrm{rot}\approx 4$ days (Table \ref{tab:targets_properties}).
    
    \item Since we used the GMRT for our observations, the declinations of the stars must be North of -53$^\circ$.
\end{enumerate}

\subsection{Observation Strategy}\label{subsec:strategy}
The stars were observed at different epochs over the frequency range of 0.6--0.8 GHz with slightly different strategies. In every case, we observed during a range of rotational phases bracketing at least one of the magnetic nulls.
The motivation behind this strategy comes from the theoretical prediction that the radio pulses due to ECME will be visible around the magnetic null phases for a star with an axi-symmetric dipolar magnetic field \citep[e.g.][]{leto2016}. 
Indeed for the MRPs HD\,133880 and HD\,35298, the ECME pulses are observed around their magnetic nulls \citep{das2018,das2019b} despite the fact that neither of them has a purely axi-symmetric dipolar magnetic field \citep{kochukhov2017,shultz2018}. However, for the rest of the MRPs, significant offsets \citep[as large as 0.1 rotational phases, ][]{leto2020a} between the rotational phases of pulse arrival and the nearest magnetic null phase are observed. The possible reasons behind such offsets include incorrect ephemerides, complex surface magnetic fields, and propagation effects in the stellar magnetosphere.

We originally observed the stars only over a narrow rotational phase window ($\pm 0.03$ rotation cycles). However following reports of offsets larger than this window, we increased the width of the rotational phase window around the magnetic nulls up to $\pm 0.35$ rotation cycles. In our sample, one star (HD\,12447) was observed for nearly one full rotation cycle. Among the remaining stars, two were observed over a rotational phase window of width $\leq \pm 0.03$ cycles around a magnetic null.


\begin{deluxetable*}{rccllrcclr}
\scriptsize
\centering
\tablecaption{Physical properties of the eight MRPs reported here. $T_\mathrm{eff}$ is the stellar effective temperature and $B_\mathrm{d}$ is the surface magnetic dipole strength. $i$ and $\beta$ correspond to the inclination angle (angle between the stellar rotation axis and the line of sight) and the obliquity (angle between the rotation and dipole axes), respectively. Also given are the reference Heliocentric Julian Day $\mathrm{HJD_0}$ used to calculate the rotational phases. The values in parentheses represent symmetric error bars. \label{tab:targets_properties}}  
\tablehead{
\hline
HD  & Mass & Radius & $T_\mathrm{eff}$ & $B_\mathrm{d}$ &  Distance & $i$ & $\beta$ & \multicolumn{2}{c}{Ephemeris} \\
  & ($M_\odot$) & $R_\odot$ & (kK) & (kG)  &  (parsec) & ($^\circ$) & ($^\circ$)  & $\mathrm{HJD_0}$ & $P_\mathrm{rot}$(d)
}
\startdata
\hline
12447 &  $2.6^{+0.2}_{-0.3}$\tablenotemark{\scriptsize a} & 2.7(0.4)\tablenotemark{\scriptsize a} & 10.0(0.7)\tablenotemark{\scriptsize a} & $2.4^{+0.7}_{-0.5}$\tablenotemark{\scriptsize b} & 50(2)\tablenotemark{\scriptsize c} & $38^{+13}_{-9}$\tablenotemark{\scriptsize b} & $86^{+3}_{-4}$\tablenotemark{\scriptsize b} & 2443118.328996\tablenotemark{\scriptsize *} & 1.490975(9)\tablenotemark{\scriptsize *}\\
19832 &  3.4(0.2)\tablenotemark{\scriptsize d} & 2.3(0.3)\tablenotemark{\scriptsize d} & 12.8(0.4)\tablenotemark{\scriptsize e} & $2.7^{+0.6}_{-0.3}$\tablenotemark{\scriptsize d} &  124(3)\tablenotemark{\scriptsize c} & $55^{+8}_{-6}$\tablenotemark{\scriptsize d} & $89^{+1}_{-5}$\tablenotemark{\scriptsize d} & 2442625.59(9)\tablenotemark{\scriptsize d} & 0.72776(1)\tablenotemark{\scriptsize d}\\
37017 &  8.4(0.4)\tablenotemark{\scriptsize f} & 3.6(0.1)\tablenotemark{\scriptsize f} & 21(2)\tablenotemark{\scriptsize d} & 6.2(0.9)\tablenotemark{\scriptsize f} & 378(11)\tablenotemark{\scriptsize g} & 38(2)\tablenotemark{\scriptsize f} & 57(2)\tablenotemark{\scriptsize f}  & 2443441.20(9)\tablenotemark{\scriptsize h} & 0.901186(2)\tablenotemark{\scriptsize h}\\
45583 &  3.2(0.1)\tablenotemark{\scriptsize d} & 2.12(0.06)\tablenotemark{\scriptsize d} & 13.3(0.3)\tablenotemark{\scriptsize j} & 9.1(0.3)\tablenotemark{\scriptsize d} &  326(7)\tablenotemark{\scriptsize c} & 48(2)\tablenotemark{\scriptsize d} & 70(2)\tablenotemark{\scriptsize d} & 2455521.75(6)\tablenotemark{\scriptsize d} & 1.17705(1)\tablenotemark{\scriptsize d}\\
79158 &  4.0(0.2)\tablenotemark{\scriptsize k} & 3.4(0.7)\tablenotemark{\scriptsize k} & 13.3(0.3)\tablenotemark{\scriptsize k} & 3.6(0.4)\tablenotemark{\scriptsize k} &  175(6)\tablenotemark{\scriptsize c} & 60(10)\tablenotemark{\scriptsize l} & 86(2)\tablenotemark{\scriptsize **} & 2443000.45(3)\tablenotemark{\scriptsize l} & 3.83476(4)\tablenotemark{\scriptsize l}\\
145501C & 4.0(0.2)\tablenotemark{\scriptsize d} & 2.26(0.06)\tablenotemark{\scriptsize d} & 14.5(0.5)\tablenotemark{\scriptsize m} & 5.8(0.3)\tablenotemark{\scriptsize d} &  141(1)\tablenotemark{\scriptsize c} & 49(3)\tablenotemark{\scriptsize d} & 89(1)\tablenotemark{\scriptsize d} & 2444774.98(9)\tablenotemark{\scriptsize d} & 1.02648(1)\tablenotemark{\scriptsize d}\\
170000&  $3.56^{+0.04}_{-0.49}$\tablenotemark{\scriptsize a} & 3.7(0.1)\tablenotemark{\scriptsize a} & 11.6(0.1)\tablenotemark{\scriptsize a} & $1.8^{+0.1}_{-0.2}$\tablenotemark{\scriptsize b} &  93(3)\tablenotemark{\scriptsize n} & $48^{+5}_{-4}$\tablenotemark{\scriptsize b} & $70^{+4}_{-5}$\tablenotemark{\scriptsize b} & 2442632.30626\tablenotemark{\scriptsize b} & 1.71649(2)\tablenotemark{\scriptsize p}\\
176582&  5.6(0.3)\tablenotemark{\scriptsize f} & 3.21(0.06)\tablenotemark{\scriptsize f} & 17.6(0.4)\tablenotemark{\scriptsize d} & 5.4(0.2)\tablenotemark{\scriptsize f} &  301(4)\tablenotemark{\scriptsize c} & 84(2)\tablenotemark{\scriptsize f} & $89.3^{+0.6}_{-1.4}$\tablenotemark{\scriptsize f} & 2454496.694(2)\tablenotemark{\scriptsize h} &1.581984(3)\tablenotemark{\scriptsize h}\\
\enddata
\tablenotemark{\scriptsize a}{\scriptsize \citet{sikora2019a}}
\tablenotemark{\scriptsize b}{\scriptsize \citet{sikora2019b}}
\tablenotemark{\scriptsize c}{\scriptsize \citet{gaia2018}}
\tablenotemark{\scriptsize d}{\scriptsize \citet{shultz2020}}
\tablenotemark{\scriptsize e}{\scriptsize \citet{netopil2008}}
\tablenotemark{\scriptsize f}{\scriptsize \citet{shultz2019b}}
\tablenotemark{\scriptsize g}{\scriptsize \citet{kounkel2017}}
\tablenotemark{\scriptsize h}{\scriptsize \citet{shultz2018}}
\tablenotemark{\scriptsize j}{\scriptsize \citet{semenko2008}}
\tablenotemark{\scriptsize k}{\scriptsize \citet{wade2006}}
\tablenotemark{\scriptsize l}{\scriptsize \citet{oksala2018}}
\tablenotemark{\scriptsize m}{\scriptsize \citet{netopil2017}}
\tablenotemark{\scriptsize n}{\scriptsize \citet{vanleeuwen2007}}
\tablenotemark{\scriptsize p}{\scriptsize \citet{musielok1986}}
\tablenotetext{\scriptsize*}{\scriptsize Sikora et al. in prep.}
\tablenotetext{\scriptsize**}{\scriptsize This work, see \S\ref{subsec:hd79158}}
\end{deluxetable*}

\begin{deluxetable*}{cccccll}
\footnotesize
\centering
\tablecaption{Observation details of our sample of stars. HJD stands for Heliocentric Julian Day.\label{tab:targets_obs}}  
\tablehead{
\hline
HD & Date & HJD range & Band & Eff. band & Flux & Phase \\
 & of Obs. & $-2.45\times 10^6$ & $\Delta\nu$ (MHz) & $\Delta\nu_\mathrm{eff}$ (MHz) & calibrator & calibrator
}
\startdata
\hline\hline
12447 & 2018--12--28 & $8481.10\pm0.15$ & 550--950 & 570--804 & 3C48 & J0204+152\\
 & 2019--11--10 & $8798.25\pm0.18$ & 550--950 & 570--804 & 3C48 & J0204+152\\
 & 2019--12--14 & $8832.07\pm0.04$ & 550--950 & 560--814 & 3C48 & J0204+152\\
 & 2020--01--02 & $8851.23\pm0.03$ & 550--950 & 570--804 & 3C48, 3C147 & J0204+152\\
 & 2020--01--03 & $8852.14\pm0.16$ & 550--950 & 570--804 & 3C48, 3C147 & J0204+152\\
 & 2020--01--04 & $8853.14\pm0.16$ & 550--950 & 570--804 & 3C48, 3C147 & J0204+152\\
 & 2020--01--05 & $8854.14\pm0.15$ & 550--950 & 570--804 & 3C48, 3C147 & J0204+152\\
\hline
19832 & 2021--03--27 & $9300.86\pm0.10$ & 550--950 & 570--804 & 3C48 & J0318+164 \\
\hline
37017 & 2018--05--11 & $8249.85\pm0.03$ & 550--750 & 570--667 & 3C48 & J0607--085\\
 & 2018--05--12 & $8250.93\pm0.03$ & 550--750 & 560--726 & 3C147 & J0607--085\\
\hline
45583 & 2020--12--01 & $9185.35\pm0.12$ & 550--950 & 570--804 & 3C286, 3C48 & J0607--085  \\
 & 2020--12--09  & $9193.34\pm0.13$ & 550--950 & 570--804 & 3C286, 3C48 & J0607--085\\
\hline 
79158 & 2019--11--26 & $8813.51\pm 0.21$ & 550--950 & 570--804 & 3C48, 3C286 & J0834+555\\
 & 2020--01--31 & $8880.31\pm0.21$ & 550--950 & 570--804 & 3C48, 3C286 & J0834+555 \\
 & 2021--02--10 & $9256.30\pm 0.22$ & 550--950 & 570--804 & 3C48, 3C286 & J0834+555 \\
 & 2021--02--12 & $9258.28\pm0.22$ & 550--950 & 570--804 & 3C48, 3C286 & J0834+555, J1006+349 \\
 & 2021--02--27 & $9273.25\pm0.22$ & 550--950 & 570--804 & 3C48, 3C286 & J0834+555, J1006+349 \\
 & 2021--03--20 & $9294.20\pm0.23$ & 550--950 & 570--804 & 3C48, 3C286 & J0834+555, J1006+349 \\
\hline
145501C & 2021--03--02 & $9275.56\pm 0.08$ & 550--950 & 570--804 & 3C286 & J1626--298 \\
 & 2021--03--07 & $9280.53\pm 0.15 $ & 550--950 & 570--804 & 3C286 & J1517--243, J1626--298, J1714--252 \\
 & 2021--03--17 & $9290.49\pm0.16$ & 550--950 & 570--804 & 3C286, 3C48 & J1517--243, J1626--298\\ 
\hline
170000 & 2018--12--17 & $8469.84\pm0.17$ & 550--950 & 570--804 & 3C48, 3C286 & J1634+627\\
\hline
176582 & 2015--07--27 & $7231.24\pm0.03$ & 591--624 & 594--620 & 3C286 & J1924+334 \\
   &  2018--06--03 & $8272.54\pm0.05$ & 550--750 & 565--726 & 3C48 & J2015+371\\
\enddata
\end{deluxetable*}

The data were acquired using the GMRT over the years 2015--2021. The earliest data (year 2015) were acquired for the star HD\,176582, prior to the upgrade of the observatory. These data span the frequency range 591--624 MHz, divided into 256 channels. The time resolution was 16 seconds. The rest of the data were acquired over the years 2018--2021 using band 4 (550--950 MHz) of the upgraded GMRT (uGMRT) and have different observation settings. The data acquired during the first half of the year 2018 cover the frequency range of 550--750 MHz, whereas the latter data cover the frequency range of 550--950 MHz. This change in observation setting was a result of the then-ongoing upgrade of the GMRT. Nevertheless, the effective bandwidths (the bandwidth after removing the edges with very low gain) are comparable for the data taken at different epochs. All the uGMRT data have time resolutions of 8 seconds. Table \ref{tab:targets_obs} details the times of observation, frequency ranges, and the calibrators used in each set of observations.

The data were analyzed using the Common Astronomy Software Applications \citep[\textsc{casa},][]{mcmullin2007} following the procedure described by \citet{das2019a,das2019b}.

\section{New stars displaying ECME signatures}\label{sec:results}
The lightcurves obtained for the different stars are discussed in the subsequent subsections. For each star, we evaluated the rotational phases using the following equation:

\begin{align}
    \mathrm{HJD}&=\mathrm{HJD_0}+P_\mathrm{rot}\cdot E\label{eq:ephemeris}
\end{align}

The reference heliocentric julian day $\mathrm{HJD_0}$ and rotation period $P_\mathrm{rot}$ of each star are given in Table \ref{tab:targets_properties}. In order to identify the magnetic nulls, the rotational modulation of the stellar longitudinal magnetic field \bz~is fitted with a function of the following form:

\begin{align}
    \langle B_\mathrm{z}\rangle&=\sum_{n=0}^{N} B_n\sin(2\pi n\phi_\mathrm{rot}+\phi_n)\label{eq:bz_curve}
\end{align}

Where $\phi_\mathrm{rot}$ is the rotational phase and $N$ is an integer (chosen based on the reduced $\chi^2$).

\subsection{HD\,12447}\label{subsec:hd12447}
\begin{figure}
    \centering
    \includegraphics[width=0.48\textwidth]{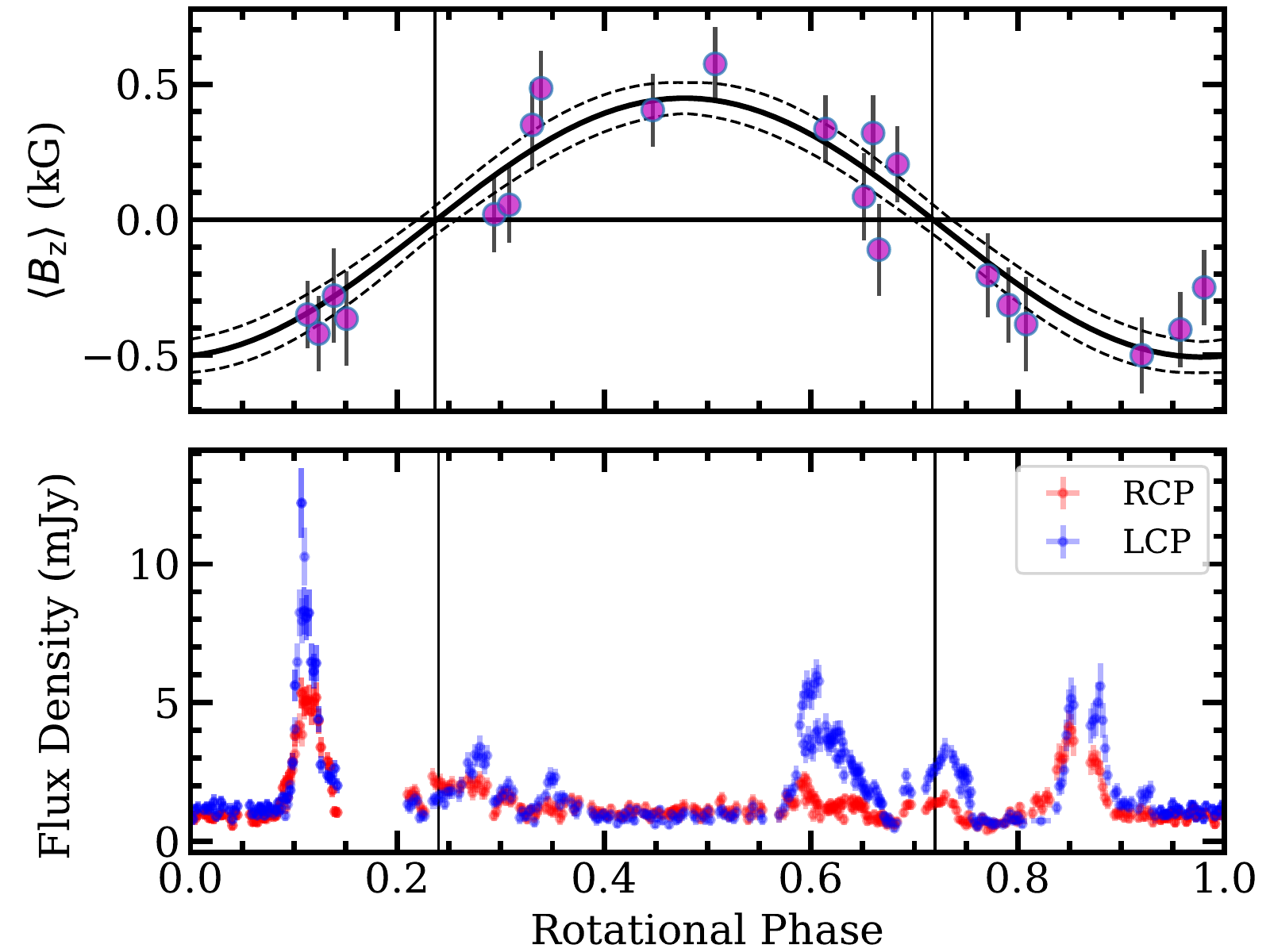}
    \caption{\textbf{Top:} The rotational modulation of the \bz~of the star HD\,12447. The data were reported by \citet{borra1980}. The solid curve represents a pure sinusoidal fit and the dashed curves represent the uncertainty associated with this fit. The vertical lines mark the magnetic nulls. \textbf{Bottom:} The lightcurves of the star at 0.6--0.8 GHz for the two circular polarizations. }
    \label{fig:hd12447_lightcurves_with_bz}
\end{figure}

HD\,12447 is the coolest star in our sample, with an effective temperature of 10 kK \citep{sikora2019a}. This is also the only star in our sample that was observed for one full rotation cycle in band 4 of the uGMRT. The rotational and magnetic properties of the star were reported by \citet{borra1980}. The rotation period reported there was 1.4907 days. This was refined further by adding newer measurements of \bz, yielding a rotation period of 1.490975(9) days (Sikora et al. in prep.). This rotation period was used to phase both magnetic and radio data (Table \ref{tab:targets_properties}).
We used a sinusoidal function ($N=1$ in Eq. \ref{eq:bz_curve}) to model the rotational modulation of \bz~(top panel of Figure \ref{fig:hd12447_lightcurves_with_bz}). The rotational phases corresponding to the magnetic nulls are $0.24\pm0.02$ and $0.72\pm0.02$.

As mentioned already, we observed the star for nearly one full rotation cycle. The lightcurves in LCP and RCP are shown in the bottom panel of Figure \ref{fig:hd12447_lightcurves_with_bz}. We find that the star shows significant variability throughout its rotation cycle. We also find the variation to be extremely stable (since we had overlap in rotational phase ranges covered on different days). This suggests that the underlying phenomenon giving rise to such enhancements are highly stable, both spatially and temporally. 
The observed variation with rotational phase is, however, quite different from the one that we expect to see due to ECME for a star with an axi-symmetric dipolar magnetic field, in the sense that the enhancements are not particularly confined to regions near the magnetic null phases. 
However, they are highly unlikely to be of gyrosynchrotron origin for the following reasons:

\begin{enumerate}
    \item The variation in the radio lightcurve is not smoothly correlated with that of \bz.
    \item The most notable feature in the lightcurve is the pulse that lies around phase 0.1 (bottom panel of Figure \ref{fig:hd12447_lightcurves_with_bz}). 
    The value of $\Delta\phi_\mathrm{rot}$ over which the LCP pulse reaches its peak from the basal flux density is only 0.02 cycles, significantly smaller than the timescale for variation of the gyrosynchrotron flux density (\S\ref{sec:MRP_signature}). Its peak flux density is nearly an order of magnitude higher than the basal flux density. 
    Gyrosynchrotron emission, on the other hand, is not known to vary by such a large amount (an order of magnitude) with stellar rotational phase. The FWHM of this pulse under consideration is $\approx 0.02$ cycles. 
   The corresponding emission cone can be obtained using the following equation:
    \begin{align}
    \cos\alpha &=\cos i\cos\beta +\sin i\sin\beta\cos 2\pi(\phi_\mathrm{rot}-\phi_0)\label{eq:angle_los_B}
    \end{align}
    where $\alpha$ is the angle between the line-of-sight and the magnetic dipole axis at a rotational phase $\phi_\mathrm{rot}$, and $\phi_0$ corresponds to the rotational phase when the line-of-sight is closest to the North pole (maximum of \bz). Using $\phi_0$=0.477 (Figure \ref{fig:hd12447_lightcurves_with_bz}), we obtain that the emission is directed over a cone with a half-angle of only 2$^\circ$.
    This very high directivity rules out gyrosynchrotron emission completely.
\end{enumerate}

Based on these arguments, we attribute the pulse seen at rotational phase 0.1 to ECME.
In that case, the pulse is offset from its `expected' rotational phase of arrival, i.e. the nearest magnetic null phase, by $\approx 0.1$ cycle (equivalently, $\approx70^\circ$ deflection from the magnetic dipole axis). Such offsets are however known to be common among MRPs (\S\ref{sec:radio_data}).

It is likely that the remaining weaker enhancements seen in the radio lightcurves at 570--800 MHz are also due to ECME. 
The FWHM of the enhancements observed around phases 0.6, 0.7 and 0.9 correspond to emission over cones with half angles $6^\circ$, $7^\circ$ and $3^\circ$ respectively, and respectively directed at $\approx 60^\circ$, $\approx 90^\circ$ and $66^\circ$ w.r.t. the dipole axis. Gyrosynchrotron emission cannot produce such tightly beamed emission.
It has been recently reported that the prototypical MRP CU\,Vir exhibits highly unusual features, attributed to ECME, at sub-GHz frequencies which are distributed over rotational phases in a similar fashion as we observe for the case of HD\,12447 \citep{das2021}. Such behaviour could be caused by propagation effects in a magnetosphere with strong plasma density gradients \citep[][]{das2020a,das2021}. HD\,12447, being a star with large misalignment between its rotation and dipole axes \citep[$\beta\approx 86^\circ$;][]{sikora2019a}, is likely to satisfy this condition \citep{townsend2005}. Alternately, the observation of RCP and LCP enhancements over the same rotational phase ranges might be indicative of elliptically polarized emission in the extra-ordinary mode \citep[observed from UV\,Ceti,][]{lynch2017}. However our data were not acquired in the full polar mode and hence we are unable to examine the linear polarization of the pulses.

In the future, it will be important to observe the star at higher and lower radio frequencies so as to check how the pulse-profiles vary as a function of frequency. For the MRP CU\,Vir, though it has been found to exhibit peculiar ECME pulses at sub-GHz frequencies, its lightcurves at 1--3 GHz are marked by two narrow, highly circularly polarized ECME pulses. It would be interesting to examine whether HD\,12447 shows similar behaviour.

\subsection{HD\,19832}\label{subsec:hd19832}

\begin{figure}
    \centering
    \includegraphics[width=0.45\textwidth]{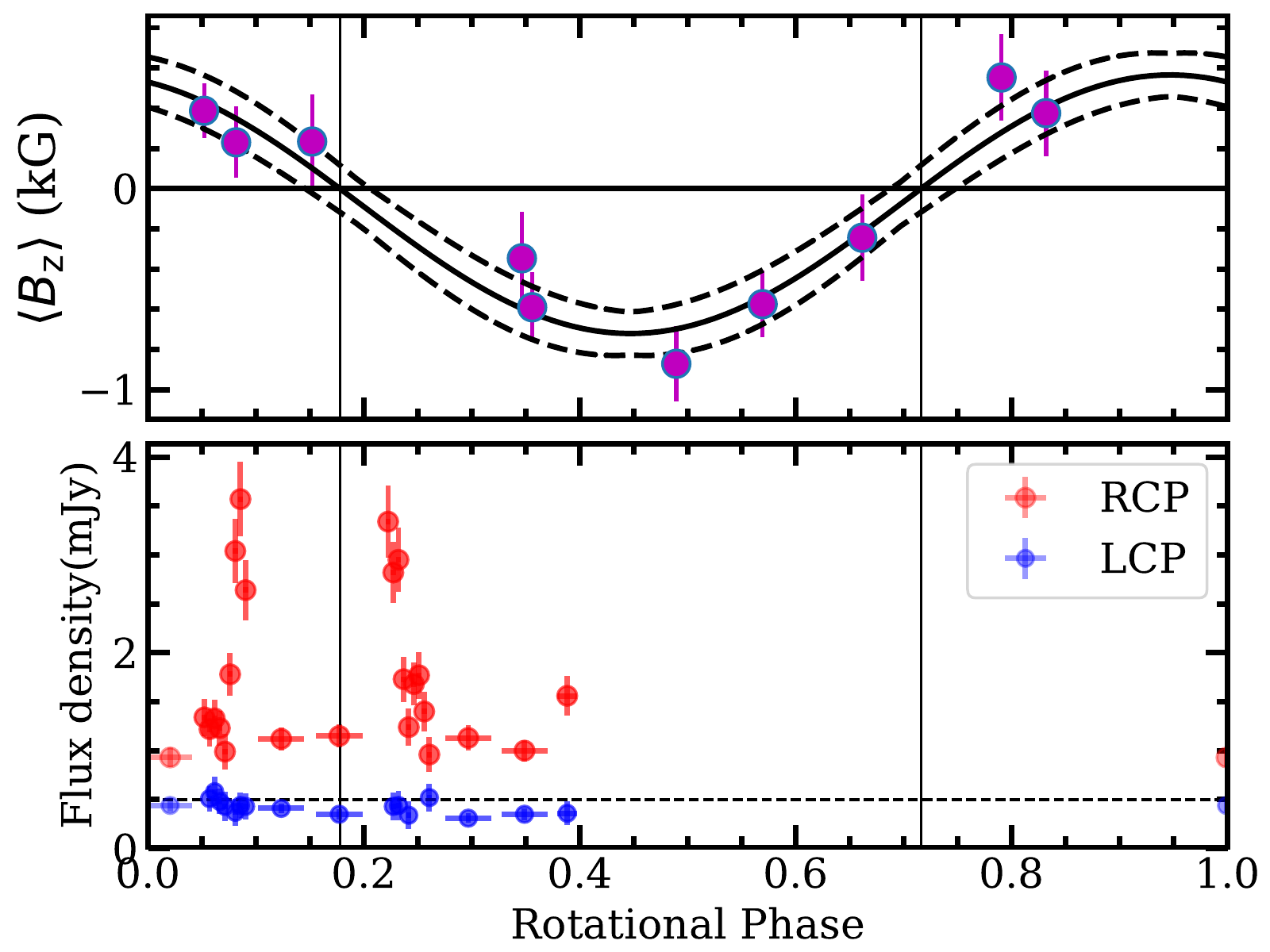}
    \caption{\textbf{Top:} The rotational modulation of \bz~for the star HD\,19832 fitted with a sinusoidal function (shown by the solid curve). These data were acquired with the Narval spectropolarimeter at the Bernard Lyot Telescope and reported by \citet{shultz2020}. The dashed curves represent the uncertainties in the fit. The vertical lines mark the magnetic null phases. \textbf{Bottom:} The lightcurves of HD\,19832 at 687 MHz. }
    \label{fig:hd19832_lightcurves_with_bz}
\end{figure}

HD\,19832 is the most rapidly rotating star in our sample. Its rotational and magnetic properties were recently reported by \citet{shultz2020}. Here we use the same ephemeris as \citet{shultz2020}. By fitting a sinusoidal function to the modulation of \bz~($N=1$ in Eq. \ref{eq:bz_curve}), we obtain the magnetic null phases to be $0.178\pm0.028$ and $0.716\pm0.028$ (top panel of Figure \ref{fig:hd19832_lightcurves_with_bz}).
We observed the star in band 4 of the uGMRT near the magnetic null at phase 0.178. The corresponding lightcurves are shown in the bottom panel of Figure \ref{fig:hd19832_lightcurves_with_bz}. As can be seen, there are significant enhancements in RCP flux density, confined to rotational phase windows of width only $\lesssim 0.06$ cycle (thus satisfying the minimum flux density gradient condition), lying on either side of the magnetic null. The maximum observed circular polarization is $78\pm5\%$. This, together with the observed sharp variation in flux density makes it a confirmed MRP. The lower limit to $T_\mathrm{B}$ is $\sim 10^{11}$ K.

The star shows a highly peculiar variation of flux density in which the `basal' flux density in RCP and LCP are significantly different. Also, there are two separate RCP pulses around the same magnetic null phase. 
It is worth noting that the star has an obliquity of $89^\circ$ \citep[Table \ref{tab:targets_properties},][]{shultz2020}. Previously, a double peaked ECME pulse was observed from the magnetic B star HD\,142990 \citep{das2019a}, which interestingly also has an obliquity close to $90^\circ$ \citep{shultz2018}. We will discuss this point in \S\ref{subsubsec:obliquity_influence}. 
The non-detection of an LCP pulse from this star could be due to the fact that the corresponding enhancement appears at a rotational phase not covered by our data. Alternately, the LCP enhancement can have very different cut-off frequencies \citep[as is the case for CU\,Vir,][]{das2021}. 
A unique feature observed for this star is the different basal flux densities (over 0.1--0.2 rotational phases) at RCP and LCP. It is possible that the narrow RCP enhancements are superposed on a much broader RCP enhancement.
Observation of this star over a full rotation cycle will be highly useful to understand its peculiar behavior.

\subsection{HD\,37017}\label{subsec:hd37017}
\begin{figure*}
    \centering
    \includegraphics[width=0.8\textwidth]{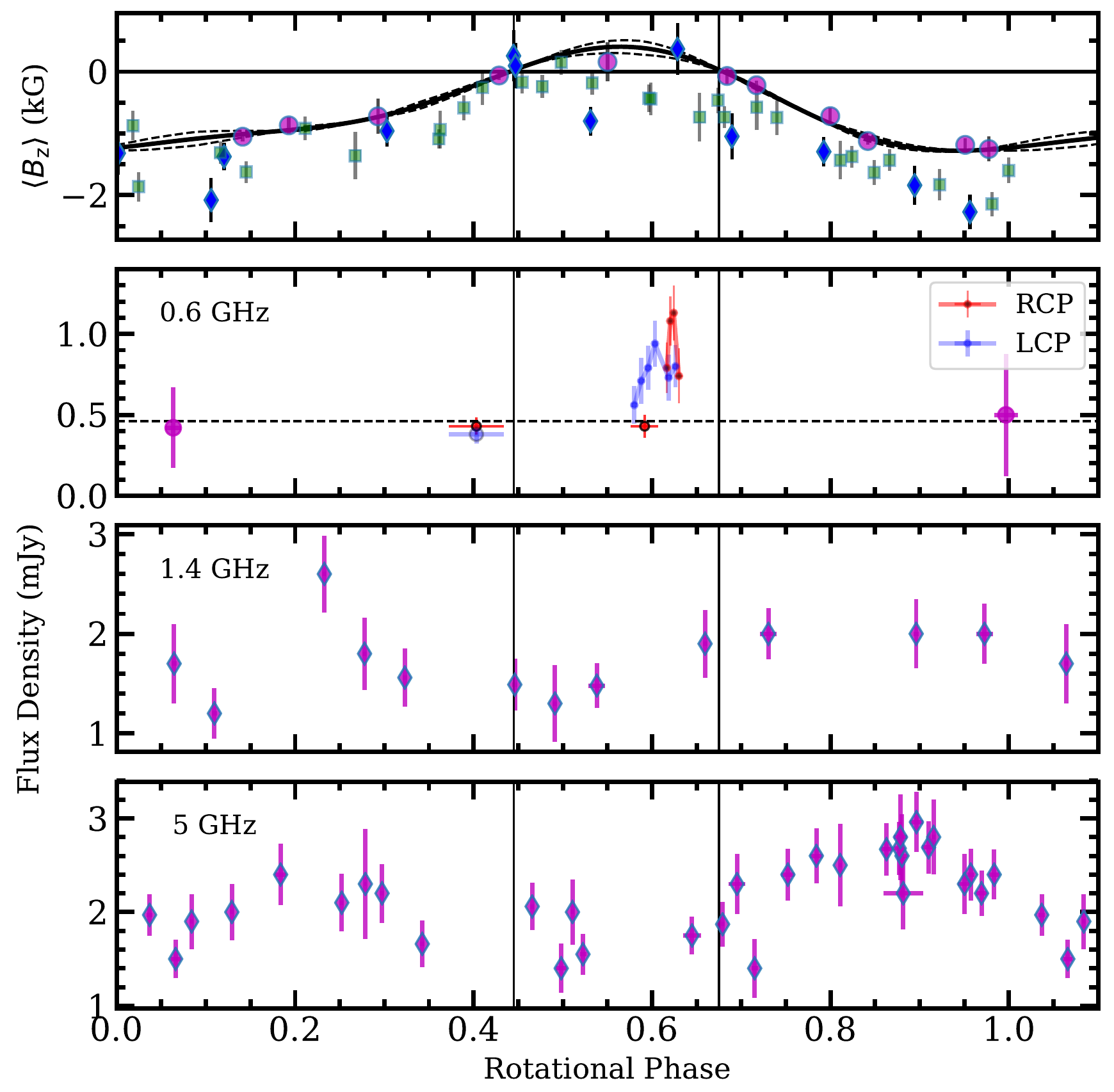}
    \caption{The variation of \bz~and radio flux densities of the star HD\,37017 with rotational phase. \textbf{Top:} The \bz~data for the star. The magenta circles represent data acquired with the ESPaDOnS spectropolarimeter at the Canada France Hawaii Telescope \citep[][see also \S\ref{sec:spectro_data_hd37017}]{shultz2018}, the blue diamonds represent data reported by \citet{borra1979}, and the green squares represent data reported by \citet{bohlender1987}. The solid curve represents a function of the form Eq. \ref{eq:bz_curve} with $N=2$ fitted with only the ESPaDOnS measurements. The dashed lines around it shows the uncertainty in the fitting. \textbf{Bottom three panels:} The flux density measurements at different radio frequencies. Except for the red and blue points in the second panel, all other measurements correspond to total intensity. The vertical lines mark the (supposed) magnetic nulls.}
    \label{fig:hd37017_combined_lightcurves_with_bz}
\end{figure*}

HD\,37017 is the most distant star in our sample (Table \ref{tab:targets_properties}). It is in a close binary system with a late B star \citep{bolton1998}. \citet{shultz2018} used historial \bz~measurements \citep{borra1979,bohlender1987} along with new spectropolarimetric data acquired with the ESPaDOnS spectropolarimeter at the Canada France Hawaii Telescope (CFHT) to obtain a rotation period of 0.901186(2) days. They also found that the rotational modulation of \bz~for this star can be well-modelled by assuming that the star has a centred dipole. This scenario however changed with the subsequent addition of new ESPaDOnS \bz~measurements (\S\ref{sec:spectro_data_hd37017}). We found that the combined ESPaDOnS data can be better modelled with the inclusion of a second harmonic to the rotational modulation of the \bz~function ($N=2$ in Eq. \ref{eq:bz_curve})\footnote{We did not include the older measurements in the fit to avoid introducing systematics due to the use of different methodologies.}. The fit along with the \bz~measurements are shown in the top panel of Figure \ref{fig:hd37017_combined_lightcurves_with_bz}. 
It can be clearly seen that the magnetic nulls and the maximum of \bz~are very closely spaced. In fact, based on the actual measurements, it is unclear whether there are indeed two magnetic nulls. At the time of scheduling the radio observations, we did not have access to the newer \bz~data that reveals the non-dipolar nature of the stellar magnetic field. We hence used the null phases predicted by the \bz~curve of \citet{shultz2018} which are 
 $0.41\pm0.02$ and $0.62\pm0.03$. Based on the solid curve in the top panel of Figure \ref{fig:hd37017_combined_lightcurves_with_bz}, the latter rotational phase is rather close to the \bz~maximum. But as mentioned already, the existence of such a (positive) \bz~maximum is not well-established.

We observed the star near both (supposed) magnetic nulls. The star was detected on both days of our observations. Over the rotational phase window 0.37--0.43 (observed on 2018 May 11), the average flux density (total intensity) is $0.44\pm 0.08\,\mathrm{mJy}$. The RCP flux density is $0.41\pm 0.07\,\mathrm{mJy}$ and the LCP flux density is $0.37\pm 0.07\,\mathrm{mJy}$, which are consistent with each other within error bars. We could not examine the time-variability of the flux density within the observation duration as the target did not present a sufficient flux density.

Near the other magnetic null, we found the target to be much brighter. The lightcurves (RCP and LCP) corresponding to this magnetic null are shown in the top panel of Figure \ref{fig:hd37017_combined_lightcurves_with_bz} (red and blue markers near rotational phase 0.6). While the maximum observed circular polarization is only $37\pm 11\%$, this is significantly more polarized than that typically observed for gyrosynchrotron emission at $\lesssim 1\,\mathrm{GHz}$ \citep[e.g.][]{leto2012,leto2017,leto2018,das2018,das2021}.
The variability of flux density with rotational phase is significant as well as confined to a narrow rotational phase range (the observation duration was only 0.06 cycles long, clearly satisfying the minimum flux density gradient condition introduced in \S\ref{sec:MRP_signature}) indicating highly directed emission. For the maximum observed flux density $\sim 1$ mJy, the lower limit to $T_\mathrm{B}$ is $\sim 10^{11}$ K. If we assume that the basal flux density of the star is $\approx 0.4\,\mathrm{mJy}$ (the flux density observed near the other magnetic null), the maximum flux density observed here corresponds to an enhancement by a factor of $\approx 3$. This makes it an MRP candidate.

In the past, radio observations of this star were reported by \citet{drake1987}, \citet{linsky1992}, \citet{leone1993} and \citet{leone2004}. These observations revealed that the star exhibits a positive spectral index ($S_\nu\propto\nu^\alpha$) of $\alpha=0.15$ over 1.4--22.5 GHz \citep{leone2004}. Also, the radio flux density at 5 GHz was found to be modulated with rotation with the maxima coinciding with the \bz~extrema and the minima coinciding with the null of \bz, which is the characteristic of gyrosynchrotron emission \citep{leone1993}. Over the full rotation cycle, the flux density was found to vary by a factor of $\approx 2$ \citep{linsky1992,leone1993}. \citet{chandra2015} reported detection of this star at 610 MHz around rotational phase 0.03 (according to the ephemeris used here). Their flux density measurement of $0.59\pm0.32\,\mathrm{mJy}$ is consistent with the flux density that we observed near rotational phase 0.4, and also the basal flux density around the null at phase 0.60. By using a 1.4 GHz measurement at a similar rotational phase, \citet{chandra2015} obtained a spectral index of $0.2\pm0.1$, consistent with the value reported by \citet{leone2004}.

In order to understand the significance of the enhancement that we observed in our data, we examined the rotational modulation due to gyrosynchrotron. For that, we reanalyzed all the archival VLA data at 5 GHz and 1.4 GHz (excluding those that do not have a suitable calibrator). These include data that have not been reported previously to the best of our knowledge, as well as those reported in the past. The resulting lightcurves are shown in the third and bottom panels of Figure \ref{fig:hd37017_combined_lightcurves_with_bz}. The second panel contains our uGMRT flux densities (red and blue markers representing RCP and LCP respectively) along with other available measurements (total intensity) at a similar frequency which were also acquired with the GMRT, but before its upgrade \citep[magenta points in the topmost panel of Figure \ref{fig:hd37017_combined_lightcurves_with_bz},][]{chandra2015}. 
In the 5 GHz lightcurve, the rotational modulation is clearly visible. This phenomenon has already been reported by \citet{leone1993}. But their rotational phase coverage was sparse and based on those data, \citet{leone1993} inferred that the modulation correlates with that of \bz, with the maxima of the gyrosynchrotron lightcurve coinciding with the \bz~minimum, and the minimum coinciding with the null of the \bz~curve ($\approx 0.5$ cycles, they assumed that the \bz~curve only touches zero and never becomes positive which might in fact be the case as mentioned already). With the addition of new data, it is now clear that rotational modulation at 5 GHz (and also at 1.4 GHz) is not as simple as had been thought before. It consists of two maxima, each lying between a magnetic null and the \bz~minimum. Over phases 0.4--0.6, the flux density appears to vary randomly around 1.6 mJy. One of the limitations of all the past measurements is that they all have significantly larger error bars as compared to the new uGMRT measurements. Also, we do not have past measurements around the magnetic null(s). However, based on the observed modulation at 1.4 and 5 GHz, it appears that none of the maxima of the gyrosynchrotron lightcurves occur near the magnetic null(s). 
In addition, the modulation in the gyrosynchrotron emission occurs over a significantly wider rotational phase window ($\approx 0.4$ of a rotation cycle), whereas we observed enhancements that are confined to a rotational phase window of width 0.06 cycle only. With decreasing frequency, the amplitude of variability in the gyrosynchrotron emission is expected to decrease \citep[e.g.][]{leto2020a,das2021}. Based on these results, we rule out gyrosynchrotron to be the cause of the enhancement seen at 550--800 MHz around phase 0.6. The possibility of free-free emission is also unlikely to cause such sharp variation with rotational phase \citep[furthermore, the mass-loss rate is too low to give rise to the observed radio emission,][]{drake1987}. Finally we rule out an origin related to binarity, since neither of the components is expected to have a strong enough wind to give rise to radio emission via wind-wind collision.
Hence the only way to produce the enhancements under consideration is via directed emission. Among the magnetic hot stars, ECME is the only known mechanism that satisfies all these requirements (directed emission, visible near the magnetic null). We therefore attribute the enhancements seen at 550--850 MHz to ECME.

To summarize, the arguments in favour of HD\,37017 being a star capable of producing ECME are as follows:

\begin{enumerate}
    \item The enhancement observed in band 4 occurs over a very narrow range of rotational phases (0.06 cycle only). Gyrosynchrotron is not known to give rise to a systematic variation in flux density over such a small phase window. Besides, the modulation seen for this star at higher radio frequencies confirms that gyrosynchrotron emission varies smoothly and gradually with rotation. 
    \item The enhancement was observed near/at a magnetic null.
    \item Based on our own measurement, the basal flux density in band 4 is $\approx 0.4$ mJy. Thus we observe enhancement by a factor of $\approx 3$. This is comparable to (but larger than) the magnitude of variation seen due to gyrosynchrotron at higher radio frequencies.
\end{enumerate}

While the star clearly exhibits ECME, the following points require clarification (or need to be examined):

\begin{enumerate}
    \item Full rotational phase coverage is necessary to clarify the reason behind the absence of any enhancement around the supposed magnetic null at phase 0.4, whether it is offset from that rotational phase, or whether it is indeed absent. The latter will support the idea that \bz~has only one magnetic null.
    \item Though we mentioned that the secondary star of the binary system is unlikely to play a role in the observed emission, it will nevertheless be important to observe the star around its magnetic nulls but at different orbital phases. As the data in band 4 were acquired on two consecutive days and the orbital period of the system is $\approx 19$ days \citep{bolton1998}, both datasets correspond to similar orbital phases.
\end{enumerate}

In the ideal case of a star with an axi-symmetric dipolar magnetic field, the magnetic null phase lies in between the RCP and LCP pulses \citep{leto2016}. In the case of HD\,37017, the rotational phase where the RCP and LCP pulses intersect is $\approx 0.62$ which coincides with one of the magnetic null phases indicated by the \bz~curve of \citet{shultz2018}. It is however to be kept in mind that for several MRPs, the mid-points between RCP and LCP pulses were found to be offset from the nearest magnetic null phases \citep[e.g.][]{leto2019,leto2020a,das2021}.

\subsection{HD\,45583}\label{subsec:hd45583}
\begin{figure}
    \centering
    \includegraphics[width=0.48\textwidth]{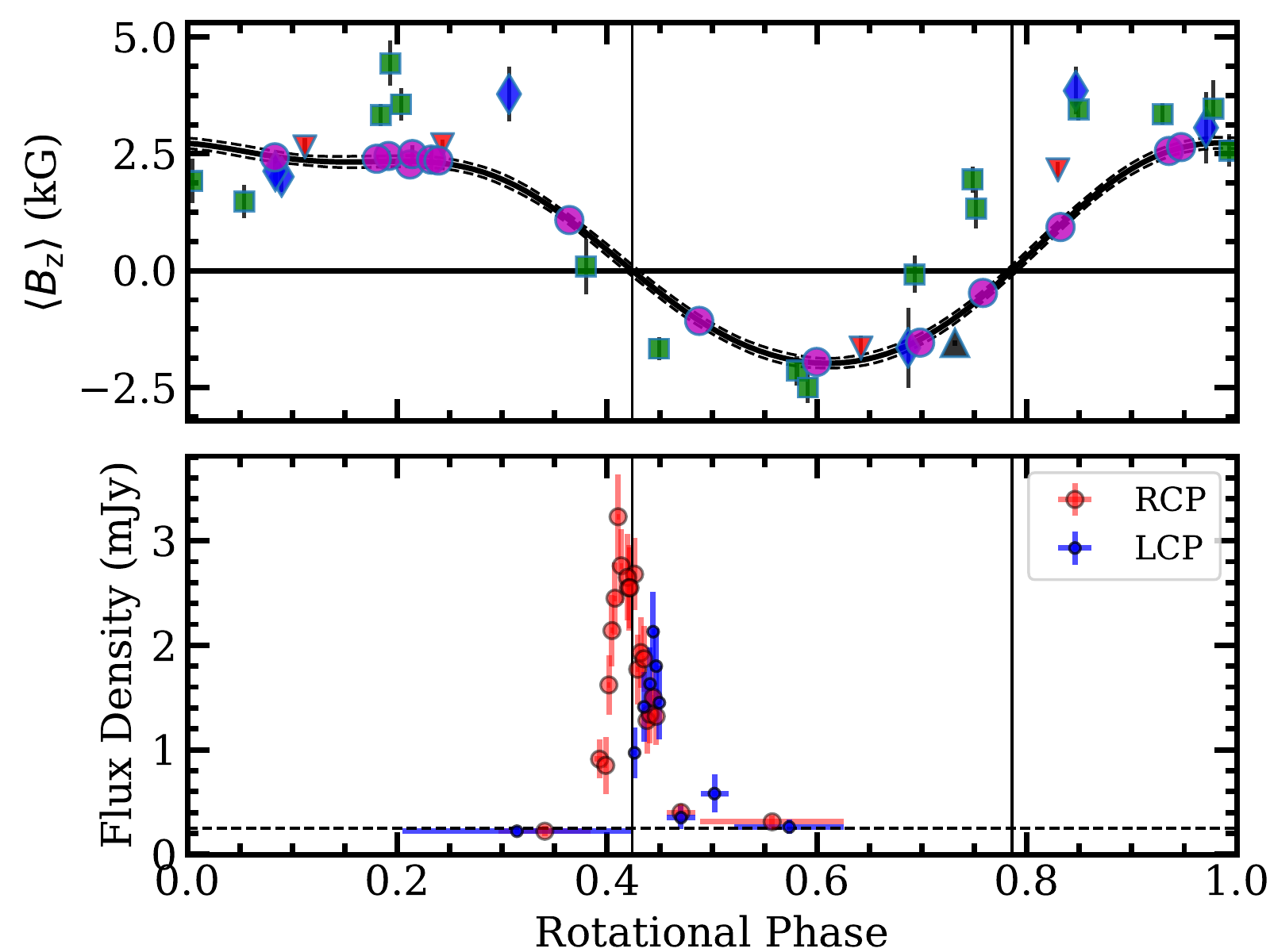}
    \caption{\textbf{Top:} The rotational modulation of \bz~for the star HD\,45583. The magenta circles, blue diamonds, the green squares, the black upper triangles and the red downward triangles correspond to \bz~data reported by \citet{shultz2020}, \citet{kudryavtsev2006}, \citet{semenko2008}, \citet{bagnulo2015} and \citet{romanyuk2017} respectively.
     The solid curve corresponds to a function of the form Eq. \ref{eq:bz_curve} with $N=3$, fitted only to the \bz~measurements from \citet{shultz2020}; the dashed lines mark the associated uncertainty. The vertical lines mark the magnetic null phases. \textbf{Bottom:} The lightcurves of HD\,45583 at 0.6--0.8 GHz. The RCP and LCP enhancements are clearly visible lying around the magnetic null at phase 0.424.}
    \label{fig:hd45583_lightcurves_bz}
\end{figure}

The magnetic properties of this star have been extensively studied \citep{kudryavtsev2006,semenko2008,bagnulo2015,romanyuk2017,shultz2020}. To locate the magnetic nulls, we use the measurements reported by \citet{shultz2020}.
Rotational modulation of \bz~exhibits clear signatures of a magnetic field more complex than a simple dipole, or a dipole+quadrupole component \citep[as first noted by][]{semenko2008}. We fitted a function of the form Eq. \ref{eq:bz_curve} with $N=3$ for \bz~vs. rotational phase (top panel of Figure \ref{fig:hd45583_lightcurves_bz}).
From the fit, we obtained the magnetic null phases to be $0.424\pm0.007$ and $0.786\pm0.005$. We observed the star around its null phase 0.424 in band 4 of the uGMRT. The lightcurves at the two circular polarizations are shown in the bottom panel of Figure \ref{fig:hd45583_lightcurves_bz}. The star exhibits clear signatures of ECME: there are enhancements in both circular polarizations; both pulses satisfy the minimum flux density gradient condition ($\Delta\phi_\mathrm{rot}<0.04$); the maximum observed circular polarization is $87\pm3\%$; and the lower limit to $T_\mathrm{B}$ is $3\times10^{12}$ K. All of these unambiguously confirm that HD\,45583 is another MRP.

\subsection{HD\,79158}\label{subsec:hd79158}
HD\,79158 (36\,Lyn) is the most slowly rotating star in our sample. Until now, the search for ECME has been limited to more rapidly rotating stars due to observational convenience of achieving rotational phase coverage with a lesser investment in telescope time. 
HD\,79158, with its well characterized rotation and magnetic properties \citep{wade2006,oksala2018}, is a well-suited starting point to extend the search to longer rotation periods.

\begin{figure}
    \centering
    \includegraphics[width=0.45\textwidth]{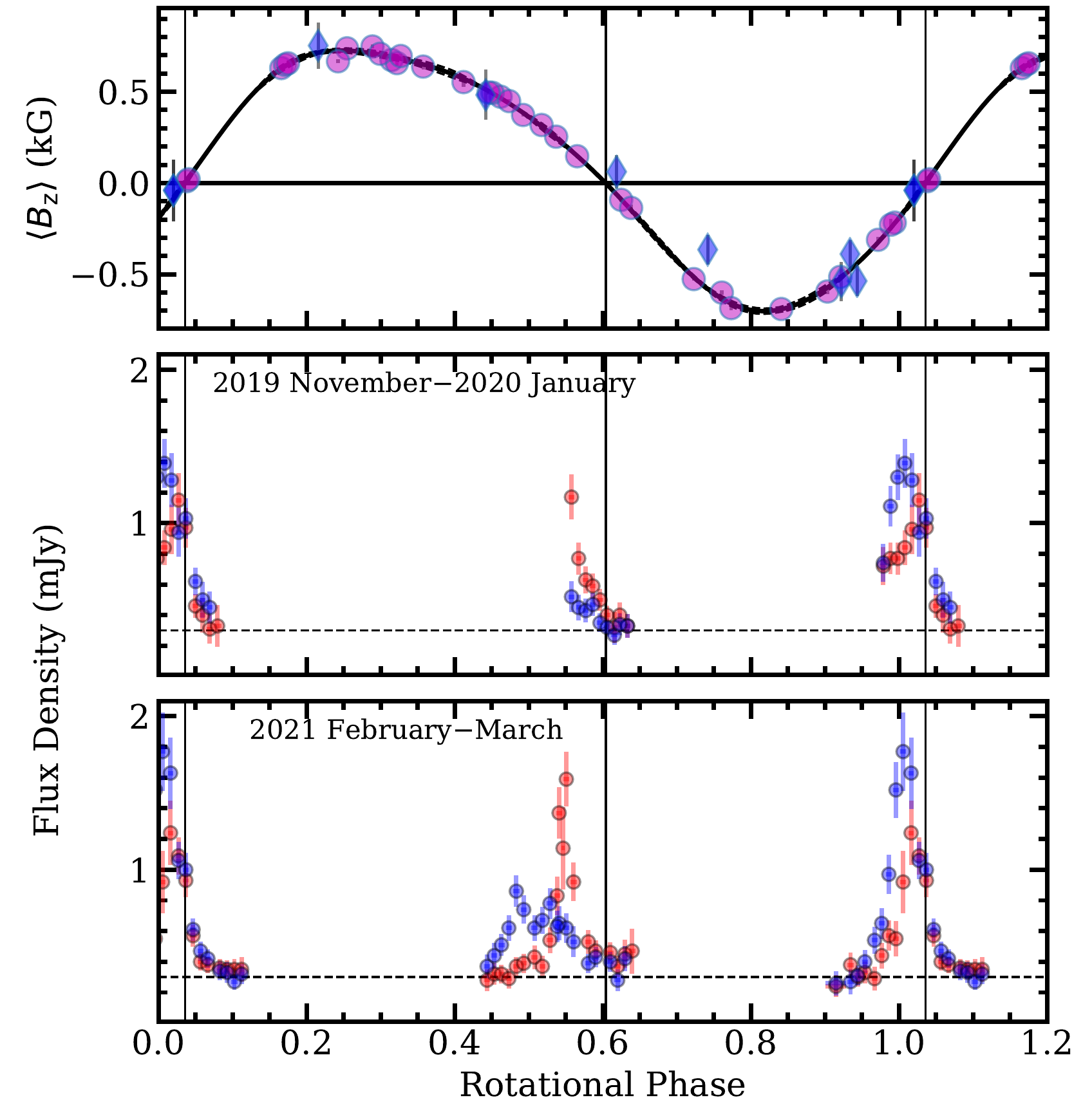}
    \caption{\textbf{Top:} The rotational modulation of \bz~for the star HD\,79158. The magenta circles and blue diamonds correspond to \bz~data reported by \citet{oksala2018} and \citet{wade2006}, respectively. The solid curve fitted to the \bz~measurements corresponds to Eq. \ref{eq:bz_curve} with $N=4$, and fitted only to the data from \citet{oksala2018}. The surrounding dashed lines show the associated uncertainty . The vertical lines mark the magnetic nulls. \textbf{Middle and bottom:} The lightcurves of HD\,79158 in band 4 near its magnetic nulls obtained at two epochs. The red and blue markers represent RCP and LCP respectively.}
    \label{fig:hd79158_lighcturves_with_bz}
\end{figure}

The obliquity for HD\,79158 in Table \ref{tab:targets_properties} was calculated using the inclination angle of \citet{oksala2018} and the method described in \citet{wade2006}, with an error propagation calculation to determine the uncertainty. However, ZDI analysis performed by \citet{oksala2018}, revealed that he magnetic field is predominantly dipolar, but has a surprisingly large (36\%) contributions from a toroidal component. As the method used to calculate the obliquity is derived presuming a simple dipole field structure, the calculation should be considered an estimate rather than a precise value.



In order to locate the nulls of \bz, we use $N=4$ in Eq. \ref{eq:bz_curve} to fit the \bz~data (reduced $\chi^2$ is 3.5). The resulting plot is shown in the top panel of Figure \ref{fig:hd79158_lighcturves_with_bz}. According to this function, the rotational phases corresponding to the magnetic nulls are $0.036\pm0.006$ and $0.604\pm0.008$. We observed near each of the two magnetic nulls at two epochs using the uGMRT in band 4. The first epoch spans 2019 November to 2020 January, and the second epoch spans 2021 February to March. From these data, we find that the star persistently produces pulses that are visible near its magnetic nulls (second and third panels of Figure \ref{fig:hd79158_lighcturves_with_bz}). 
The maximum observed circular polarization is $44\pm7\%$ and the lower limit to $T_\mathrm{B}$ is $\sim10^{11}$ K.

Figure \ref{fig:hd79158_lighcturves_with_bz} shows one interesting characteristic of the radio lightcurves, which is that the radio pulses are significantly broader (e.g. the first enhancement nearly covers 0.2 cycles) than the ECME pulses seen from other MRPs at similar frequencies. However, it has been observed that the pulse-width varies from star to star \citep[e.g.][]{das2019a,das2019b} and the width increases as we go to lower frequencies \citep[e.g.][]{das2020b}. In fact, for the star CU\,Vir, \citet{das2021} observed ECME pulses of similar width over 0.4--0.7 GHz. 
It is also interesting to find different profiles for pulses of the same polarization but visible at different magnetic nulls (e.g. see the LCP pulses in the bottom panel of Figure \ref{fig:hd79158_lighcturves_with_bz}). Such differences have been observed in the past \citep[e.g. HD\,133880;][]{das2020b} and could be due to propagation effects in a magnetosphere with an azimuthally asymmetric plasma distribution \citep{das2020a,das2020b}.

Thus, we suggest that the star HD\,79158 produces ECME since:

\begin{enumerate}
    \item Its lightcurve at our observing frequency shows persistent enhancements in flux density in both RCP and LCP near both magnetic nulls.
    \item The enhancements satisfy the minimum flux density gradient condition. The largest $\Delta\phi_\mathrm{rot}$ over which the flux density of an enhancement reaches its peak from its basal value is $\approx 0.10$ cycles.
    \item At our observing frequency (0.6--0.8 GHz), gyrosynchrotron emission is not known to give rise to an order of magnitude enhancement (from $\approx 0.2\,\mathrm{mJy}$ to 2 mJy) in flux density.
\end{enumerate}

The immediate future work on this star will be to observe it over a wider range of rotational phases and frequencies. The former is especially important as its obliquity is close to $90^\circ$ (\S\ref{subsubsec:obliquity_influence}). Wideband observation, on the other hand, will be able to clarify the point of pulse width at sub-GHz frequencies. In addition, it might be able to help us to understand how the frequency dependence of ECME pulse-width varies with stellar parameters.

\subsection{HD\,145501C}\label{subsec:hd145501c}
\begin{figure}
    \centering
    \includegraphics[width=0.48\textwidth]{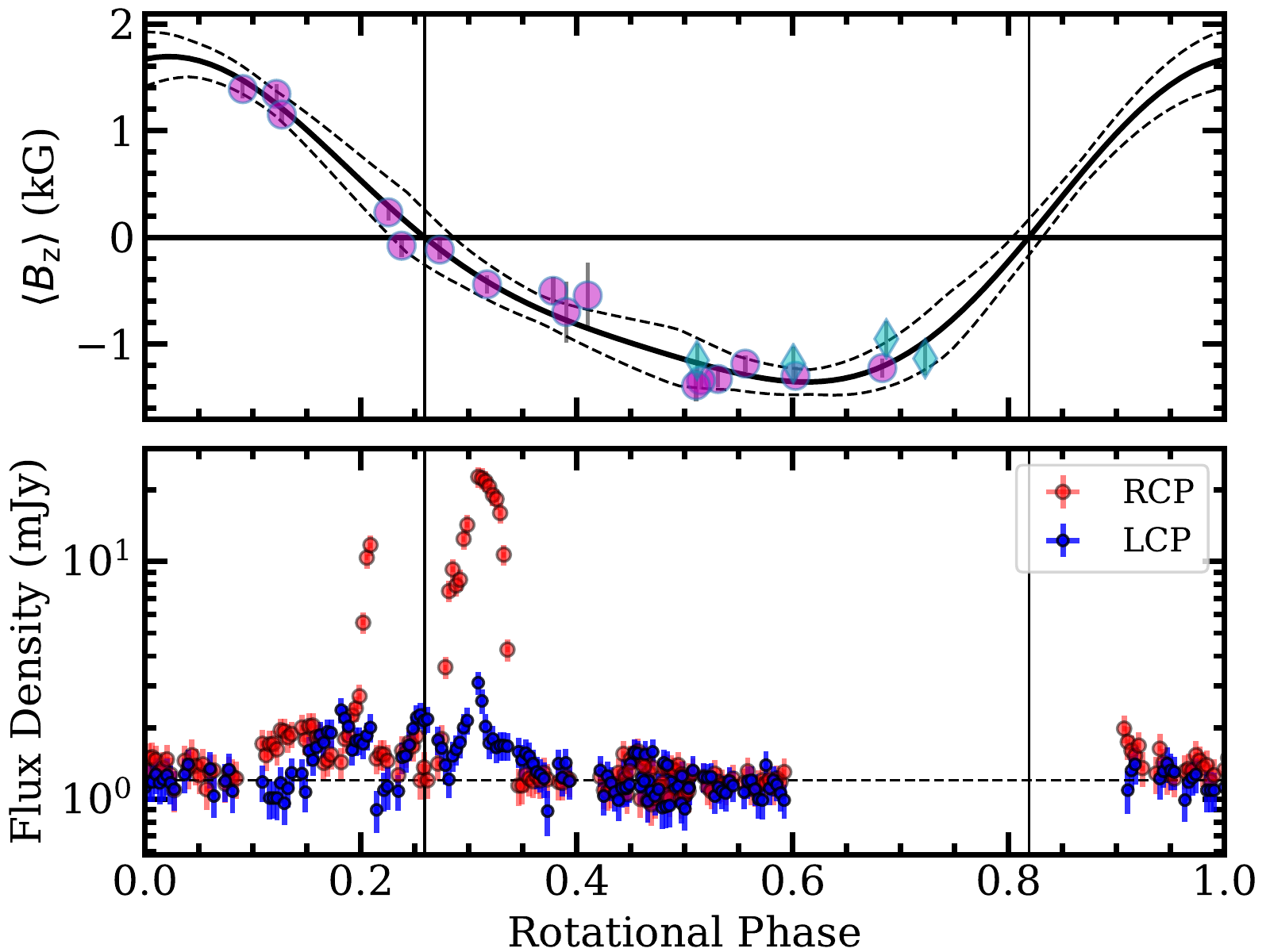}
    \caption{\textbf{Top:} The rotational modulation of \bz~for the star HD\,145501C. The measurements corresponding to the magenta points were reported by \citet{shultz2020} and those corresponding to the cyan points were reported by \citet{borra1983}. The solid curve corresponds to a function of the form given by Eq. \ref{eq:bz_curve} with $N=$ 2, and the dashed lines represent error bars associated with the fit. The vertical lines represent the magnetic nulls. As can be seen, the null at phase 0.82 is not constrained by data. \textbf{Bottom:} The radio lightcurves of HD\,145501C at 687 MHz covering the magnetic null at phase 0.26 (which is well constrained by data). Note that the Y-axis is in log scale. This is done to improve the visibility of the weaker LCP enhancements.}
    \label{fig:hd145501_lightcurves_with_bz}
\end{figure}

HD\,145501C is another magnetic B star of which rotational and magnetic properties were reported by \citet{shultz2020}. The rotational modulation of \bz~is not well-constrained, as there are not sufficient data covering the full rotation cycle (top panel of Figure \ref{fig:hd145501_lightcurves_with_bz}). Here we fit a function of the form Eq. \ref{eq:bz_curve} with $N=2$ to the variation of \bz~with rotational phase. The magnetic null phases obtained from this fit are $0.26\pm0.04$ and $0.82\pm0.03$. We however found that the value for the null at which \bz~changes from negative to positive, changes if we fit a different function to the \bz~data. This is not surprising since there are no data for \bz~around this magnetic null. The other null phase (phase 0.26 in Figure \ref{fig:hd145501_lightcurves_with_bz}), on the other hand, is well constrained. The radio data were acquired around this null (bottom panel of Figure \ref{fig:hd145501_lightcurves_with_bz}). After HD\,12447, this star has the highest fractional rotational phase coverage in our sample. As can be seen, the lightcurve in RCP is marked by two very strong enhancements around the magnetic null at phase 0.26. 
The value of $\Delta\phi_\mathrm{rot}$ over which an enhancement reaches the maximum flux density from the basal level is $\lesssim 0.04$ cycles, thus satisfying the minimum flux density gradient condition.
For the maximum observed flux density, $T_\mathrm{B}>10^{12}$ K confirming that it is a result of coherent radio emission. The maximum observed circular polarization is $76\pm3\%$. From these observations, we attribute the enhancements to ECME.

In addition to the RCP enhancements, there are also enhancements in LCP that are significantly weaker than their RCP counterparts. Such behaviour has been observed in other MRPs also \citep[e.g. HD\,133880;][]{das2020b}.

One of the most interesting observations that we made is the double RCP pulse for this star, similar to the stars HD\,19832 and HD\,142990. All three stars share one common property: they all have obliquities $\approx 90^\circ$.

\subsection{HD\,170000}\label{subsec:hd170000}

\begin{figure}
    \centering
    \includegraphics[width=0.45\textwidth]{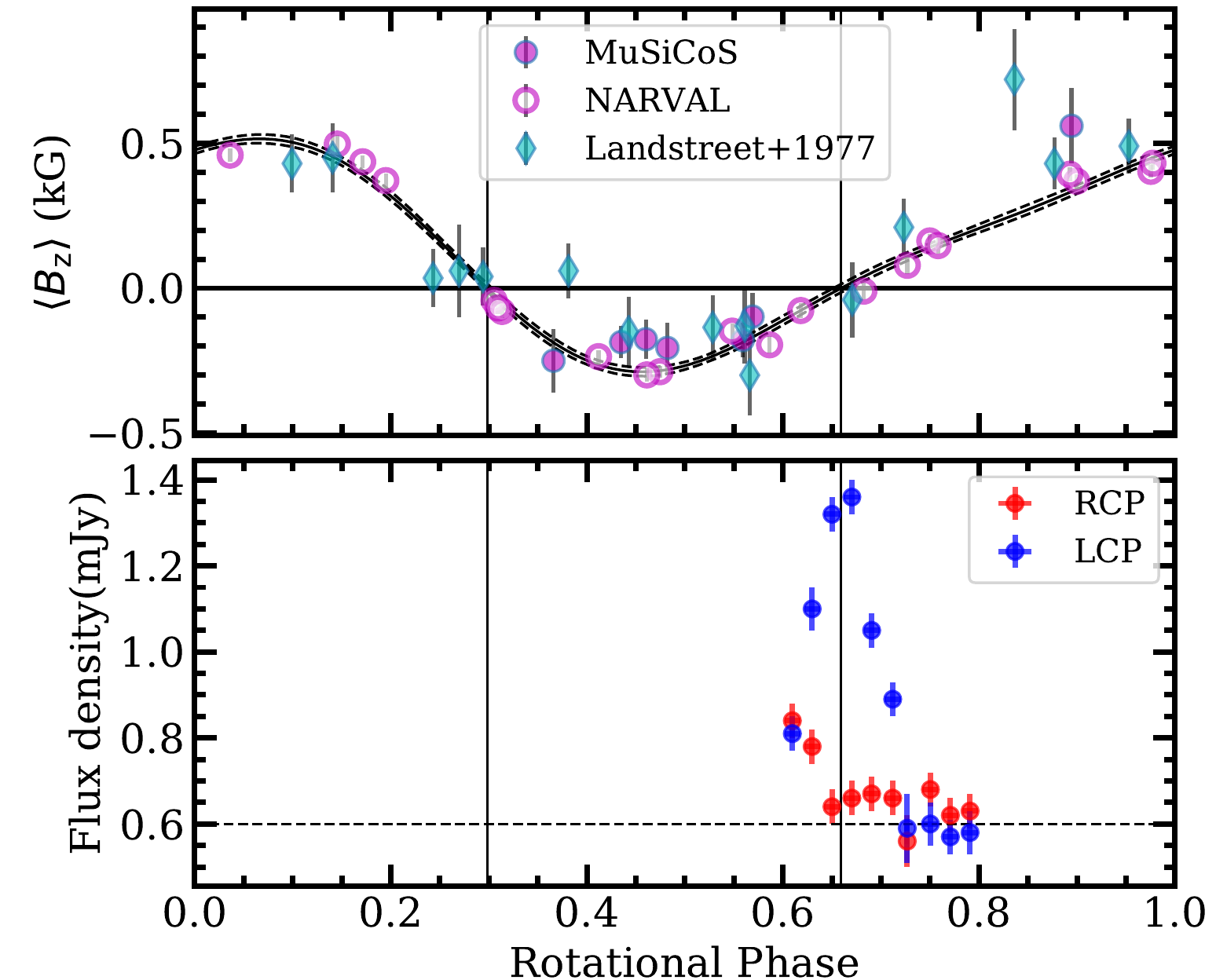}
    \caption{\textbf{Top:} The \bz~variation with rotational phase for the star HD\,170000. We fit the data with a function of the form Eq. \ref{eq:bz_curve} with $N=2$ (solid curve). The dashed lines represent the uncertainty in the fit. The magenta points represent data reported by \citet{sikora2019b} and the cyan points represent data reported by \citet{landstreet1977}. \textbf{Bottom:} The lightcurves of HD\,170000 at 687 MHz. The vertical lines represent magnetic null phases.}
    \label{fig:hd170000_lightcurves_with_bz}
\end{figure}

HD\,170000 is the second coolest star in our sample. Its rotation period was recently modified from 1.71649(2) days \citep{musielok1986} to 1.71665(9) \citep{bernhard2020}. We however find this new period to be unable to consistently phase \bz~data acquired at widely separated epochs \citep{landstreet1977, sikora2019b}. We hence chose to use the older rotation period for phasing both magnetic and radio data. The top panel of Figure \ref{fig:hd170000_lightcurves_with_bz} shows the rotational modulation of \bz~data along with the fit (of the form Eq. \ref{eq:bz_curve} with $N=2$) The magnetic null phases according to this fit are $0.298\pm0.005$ and $0.659\pm0.008$.

We observed this star around its magnetic null at phase 0.66. The lightcurves that we obtained are shown in the bottom panel of Figure \ref{fig:hd170000_lightcurves_with_bz}. We find a significant enhancement in LCP flux density with $\Delta\phi_\mathrm{rot}\approx 0.10$ cycles (thus satisfying the minimum flux density gradient condition, \S\ref{sec:MRP_signature}), right at the magnetic null phase. In addition, there is also an indication of an enhancement in RCP ahead of the start of the rotational phase window covered by our observation. The maximum observed circular polarization is $35\pm3\%$ and the lower limit to $T_\mathrm{B}$ is $\sim 10^{10}$ K. Based on the observation of a flux density enhancement confined to a rotational phase window of width $\sim 0.1$ cycles around a magnetic null phase, we suggest that the star is an MRP. In addition, circular polarization as high as $35\%$ at sub-GHz frequencies and at a magnetic null phase goes against the idea of gyrosynchrotron.

In the future, observation over a broader rotational phase window near both magnetic nulls will be highly useful to understand the properties of coherent radio emission observed from this star.

\subsection{HD\,176582}\label{subsec:hd176582}
\begin{figure}
\centering
\includegraphics[width=0.45\textwidth]{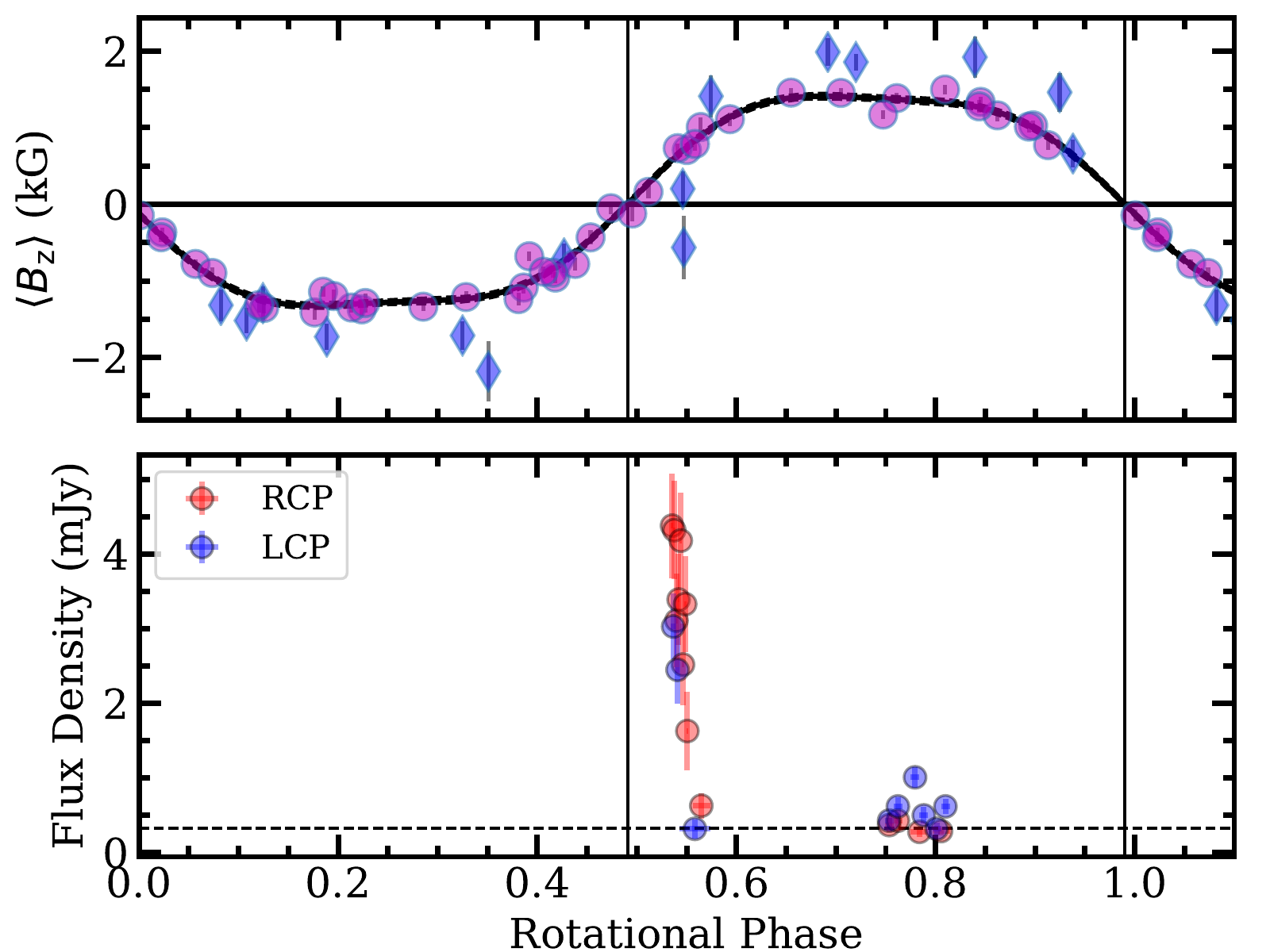}
\caption{\textbf{Top:} The rotational modulation of \bz~for the star HD\,176582. The magenta circles and blue diamonds correspond to data reported by \citet{shultz2018} and \citet{bohlender2011}, respectively. 
The vertical lines correspond to magnetic nulls. The solid curve corresponds to the function given by Eq. \ref{eq:bz_curve} with $N=3$, fitted only with the data from \citet{shultz2018}. \textbf{Bottom:} The lightcurves of HD\,176582 over 0.6--0.7 GHz. The data around the phase 0.5 were acquired using the legacy GMRT, whereas the other data were acquired using the uGMRT. 
\label{fig:hd176582}}
\end{figure}

The magnetic field in HD\,176582 and its co-rotating magnetosphere was first discovered by \citet{bohlender2011}. For this star also, \citet{shultz2018} proposed presence of higher order moments in the magnetic field. The \bz~modulation with rotational phase, fitted with a function of the form given by Eq. \ref{eq:bz_curve} with $N=3$ is shown in the top panel of Figure \ref{fig:hd176582}. According to the fitted \bz~function, the rotational phases corresponding to the magnetic nulls are $0.491\pm0.003$ and $0.990\pm0.002$. We observed the star around phase 0.5 using the legacy GMRT at 0.6 GHz in the year 2015. In addition, we observed the star around phase 0.8 using the band 4 (550--800 MHz) of the uGMRT in the year 2018. The latter rotational phase is close the rotational phase corresponding to the \bz~maximum.


In the bottom panel of Figure \ref{fig:hd176582}, we show the radio flux density measurements. 
The GMRT data clearly show a very strong enhancement near a magnetic null. The fact that the flux density observed at the null is higher by a factor of $\gtrsim 4$ than that observed around a \bz~maximum, and that the enhancement shows a sharp change in flux density over a narrow rotational phase window 
($\Delta\phi_\mathrm{rot}\approx 0.03$ cycles, satisfying the minimum flux density gradient condition),
strongly suggest that the star is also an MRP. We next calculate the lower limit to $T_\mathrm{B}$, setting the source size to be comparable to the stellar disk. For the maximum observed flux density ($\approx 4$ mJy), we find $T_\mathrm{B}>2\times 10^{12}$ K, implying that the emission is of coherent origin. This, together with the fact that the enhancement was observed near a magnetic null, confirms HD\,176582 to be an MRP.

The rotational phases corresponding to the maximum observed flux density is offset by $\approx 0.05$ cycles from the nearest magnetic null phase. However past observations have shown that such offsets are rather common among the MRPs. Offsets as large as 0.1 cycles have been reported in the rotational phases of arrival of ECME \citep{leto2020a}. Note that this star also has an obliquity close to $90^\circ$ (Table \ref{tab:targets_properties}). Unfortunately, we do not have sufficient rotational phase coverage to infer anything about the pulse-profile.
In the future, this star is worth observing over a broader rotational phase window near both magnetic nulls.

\section{Discussion}\label{sec:discussion}
With the addition of eight more magnetic hot stars to the list of known MRPs, the number of such stars has more than doubled from 7 to 15. Among the newly added stars, HD\,12447 becomes the nearest known MRP (50 pc, Table \ref{tab:targets_properties}), surpassing CU\,Vir which is at a distance of 72 pc \citep{gaia2018}. The same star (HD\,12447) is also the coolest MRP ($T_\mathrm{eff}\approx 10\,\mathrm{kK}$, Table \ref{tab:targets_properties}) known so far. This work also introduces the most slowly rotating MRP HD\,79158 (Table \ref{tab:targets_properties}). Our results therefore significantly expand the stellar parameter space of magnetic stars which can produce ECME.


In the following subsections, we discuss some of the key conclusions that we are able to draw from this work.

\subsection{The incidence of MRPs amongst magnetic hot stars}\label{subsec:ecme_rare_or_not}

\begin{figure*}
\centering
\begin{tabular}{ccc}
   \includegraphics[width=.32\textwidth]{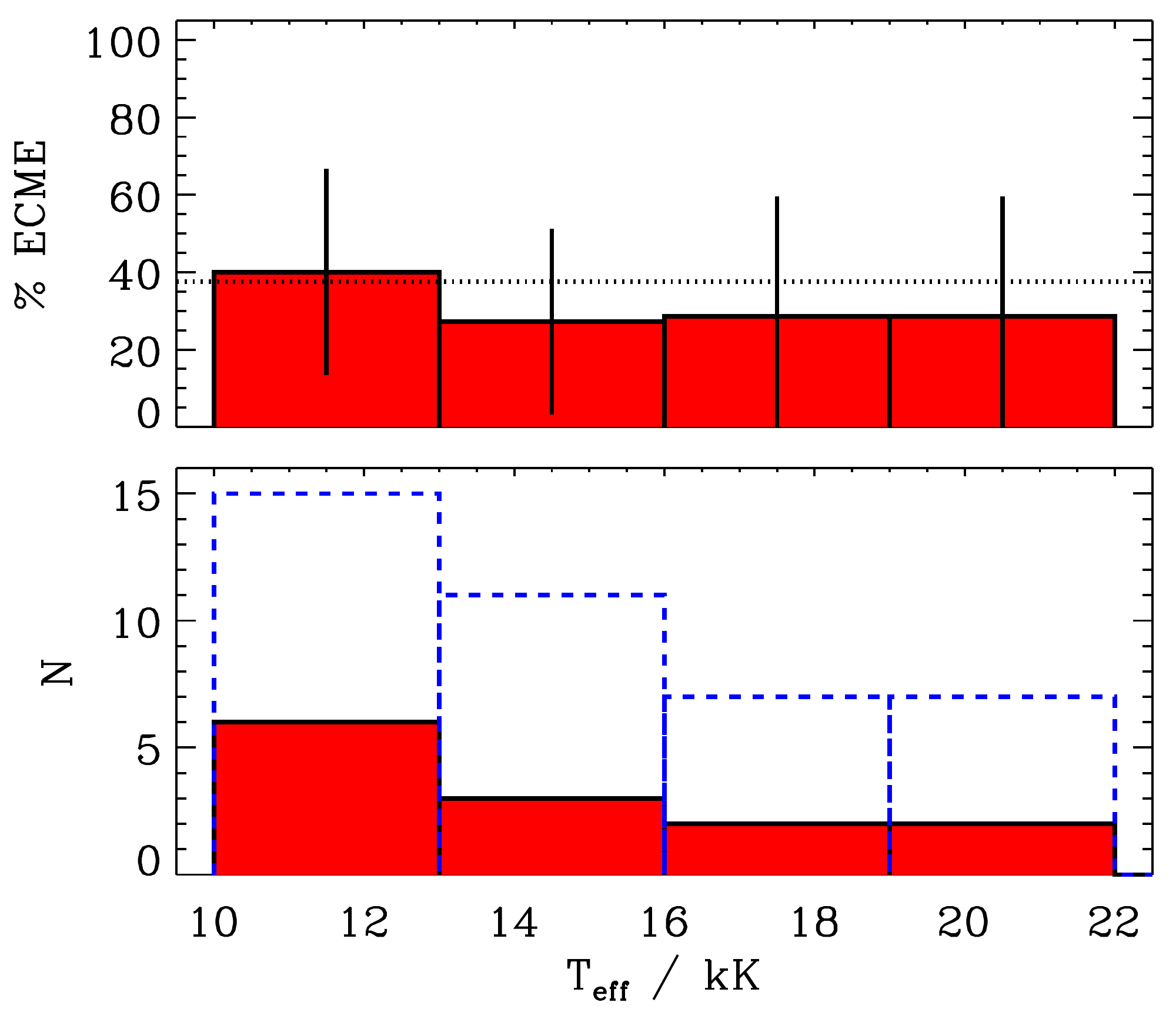} &
   \includegraphics[width=.32\textwidth]{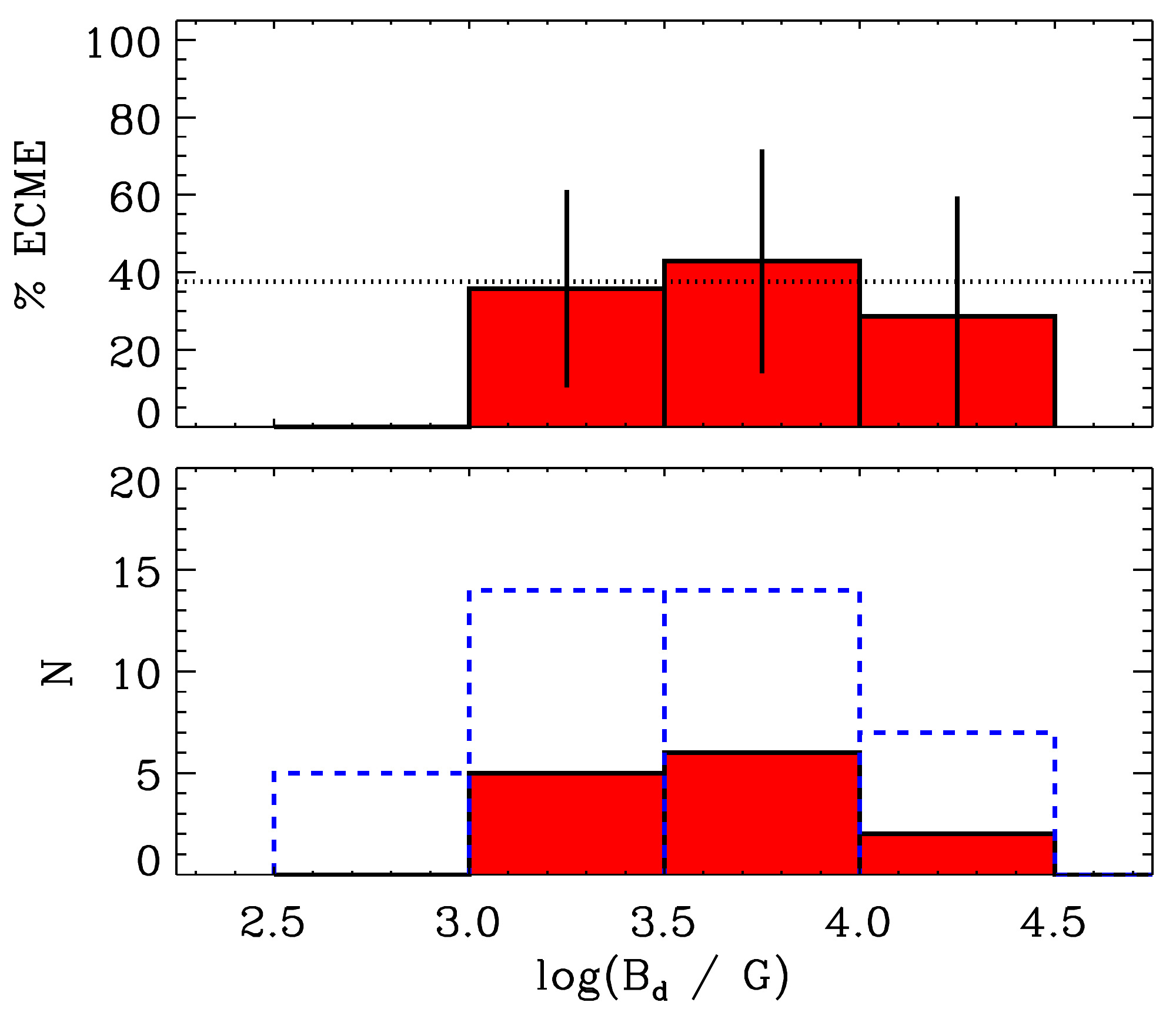} & 
   \includegraphics[width=.32\textwidth]{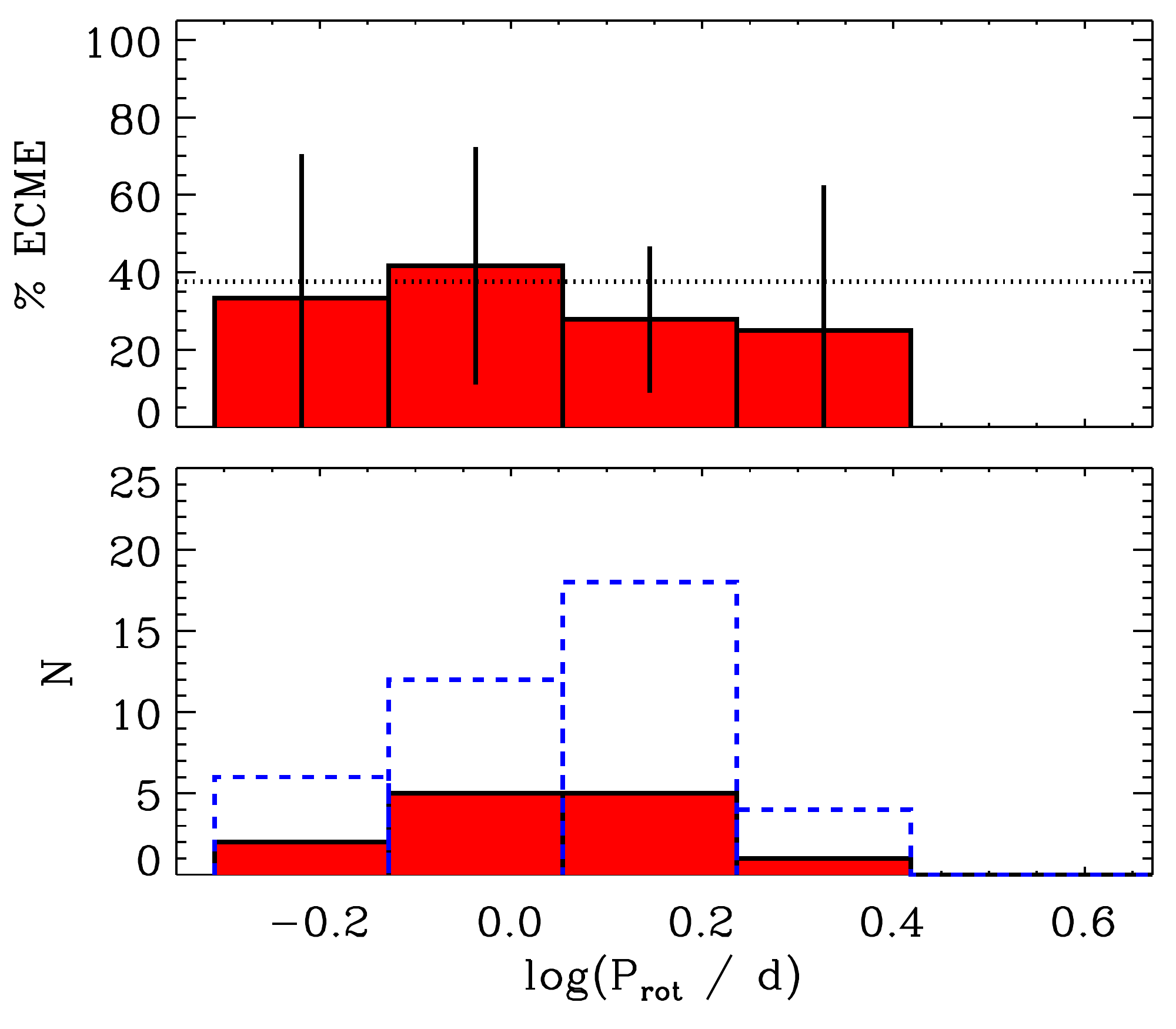} \\
\end{tabular}
\caption{Histograms of \teff~({\em left}), $B_{\rm d}$ ({\em middle}), and $P_{\rm rot}$ ({\em right}) \textit{Top:} for the percentage incidence of MRPs. The absolute numbers are shown in the bottom panels with dashed blue bars representing the number of all the comparison stars and the solid red bars representing the number of MRPs.}
\label{ecme_hist}
 \end{figure*}

The large number of ECME detections reported here, more than doubling the number of known MRPs, suggests that the phenomenon of ARE may be more common amongst magnetic early-type stars than previously supposed. In order to quantify this, we searched the literature for the known magnetic hot stars. The major studies consulted were the following: the studies of Ap, He-weak, and He-strong stars by \cite{borra1979,borra1980,borra1983} and \cite{bohlender1987,bohlender1993}; the slowly rotating Ap star study of \cite{landstreet2000}; the study of Ap stars by \cite{auriere2007}; the sample of Ap stars in open clusters presented by \cite{landstreet2007,landstreet2008}; the Herbig Ae/Be stars studied by \cite{alecian2013}; the Of?p stars examined by \cite{petit2013} and \cite{munoz2020}; the early B-type stars presented by \cite{shultz2018,shultz2019a,shultz2020}; the volume-limited sample of Ap stars conducted by \cite{sikora2019a,sikora2019b}; the samples of stars with magnetically split lines examined by \cite{mathys2017} and \cite{Chojnowski2019}; and the results of the ongoing survey at the Special Astrophysical Observatory of both field stars \citep{kudryavtsev2006,romanyuk2014,romanyuk2015,romanyuk2016,romanyuk2017,romanyuk2018,romanyuk2019,romanyuk2020} and stars in the Orion nebula \citep{romanyuk2016b,romanyuk2017b,romanyuk2019b,romanyuk2021}. The compilation of longitudinal magnetic field curves provided by \cite{bychkov2020} was also consulted, in order to include magnetic stars reported in single-star papers.

The full catalogue consists of 765 stars with at least one magnetic measurement. Comparison to this population yields an incidence fraction of just under 2\%. However, ARE can only be detected in stars with magnetic nulls; for the majority of stars in the sample only individual magnetic measurements are available, and it is therefore unknown whether these stars exhibit magnetic nulls. Furthermore, with the exception of HD\,79158, our survey was limited to stars with $P_{\rm rot} < 2$~d, whereas many magnetic stars have periods much longer than this. Finally, ARE has been detected only in stars with \teff~$\geq 10$~kK, likely because cooler stars do not possess strong enough winds to populate their magnetospheres with a sufficient electron density to generate radio emission \citep[and indeed, incoherent gyrosynchrotron has not been detected from such stars, e.g.][]{drake1987,linsky1992}. It is therefore necessary to limit the comparison sample to stars occupying the same parameter space as the stars of the survey. 

When effective temperatures were not available in the above studies, we consulted the compilations presented by \cite{kochukhov2006}, \cite{netopil2017}, and \cite{moiseeva2019}. When not available in those studies, we cross-referenced with the Str\"omgren photometric catalogue published by \cite{paunzen2015}, utilizing the {\sc idl} program {\sc uvbybeta} \citep[which implements the calibration determined by][]{napiwotzki1993}, with the calibration set by the Simbad spectral type. If Str\"omgren photometry was not available, we utilized the Johnson photometric colours obtained from Simbad together with the empirical colour tables given by \cite{worthey2011}, with reddening values found using the {\em Stilism} three-dimensional tomographic dust map \citep{lallement2014,capitanio2017, lallement2018} and distances from {\em Gaia} early Data Release 3 parallaxes \citep{gaia2021}.

In the end the catalogue contains 245 stars for which the \teff, rotation period, and ORM parameters are all available, of which MRPs comprise about $6 \pm 2$\%. However, there are only 43 stars with $9.3 < T_{\rm eff} < 23$~kK (the \teff~range in our sample when uncertainties are accounted for), $P_{\rm rot} < 2$~d, and the presence of magnetic nulls in their \bz~curves, of which MRPs comprise $32 \pm 14$\%. We note that this already high fraction is a conservative lower boundary: many of the stars have not been observed for ECME, since e.g. their magnetic fields or rotation periods were only reported within the last couple of years; further, of those that have, but in which ECME has not yet been detected, it cannot yet be ruled out that the pulses were missed due to either errors in the ephemerides, or phase offsets from the expected occurrence at the magnetic null. 

Fig.\ \ref{ecme_hist} shows histograms of \teff, $B_{\rm d}$, and $P_{\rm rot}$ for the comparison population and MRPs. The \teff~distribution of MRPs closely follows that of the comparison population, with an incidence fraction consistent with no variation with \teff. There are no MRPs with $B_{\rm d} < 1$~kG, which is likely a consequence of the choice of observing frequency (\S\ref{subsec:selection_criteria}).
Above this the incidence fraction is consistent with a flat distribution. To explore the possibility of ECME in stars with $B_\mathrm{d}<1\,\mathrm{kG}$, one will have to go to frequencies smaller than 0.7 GHz. There is some suggestion that the incidence falls off with slower rotation, however it is consistent with a flat distribution within uncertainties. We, however, would like to emphasize that the fact that all the known MRPs are relatively rapid rotators, is a result of observational convenience (see point 3 of \S\ref{subsec:selection_criteria}).
Thus within the parameter space spanned by the MRPs, there seems to be no preference for any sub-group of stellar parameters in terms of incidence fraction. This scenario might change with the discovery of more MRPs in the future.


\subsection{Comparing physical properties of the MRPs discovered}\label{subsec:compare_MRP_prop}
One of the prime motivations for continuing to search for MRPs is to have a sample large enough so as to be able to compare their physical properties and find answers to questions like what type of magnetic hot stars produce ECME. With our addition of eight MRPs, the number of MRPs known have gone up to 15. Though it cannot be called a `large' sample, we still attempt, for the first time, an investigation of the emission properties of the population and how they correlate to previously determined stellar, magnetic, and rotational parameters. Below we present the results from this exercise.

\subsubsection{Onset of ECME}\label{subsubsec:ecme_onset}
\begin{figure*}
    \centering
    \includegraphics[width=0.95\textwidth]{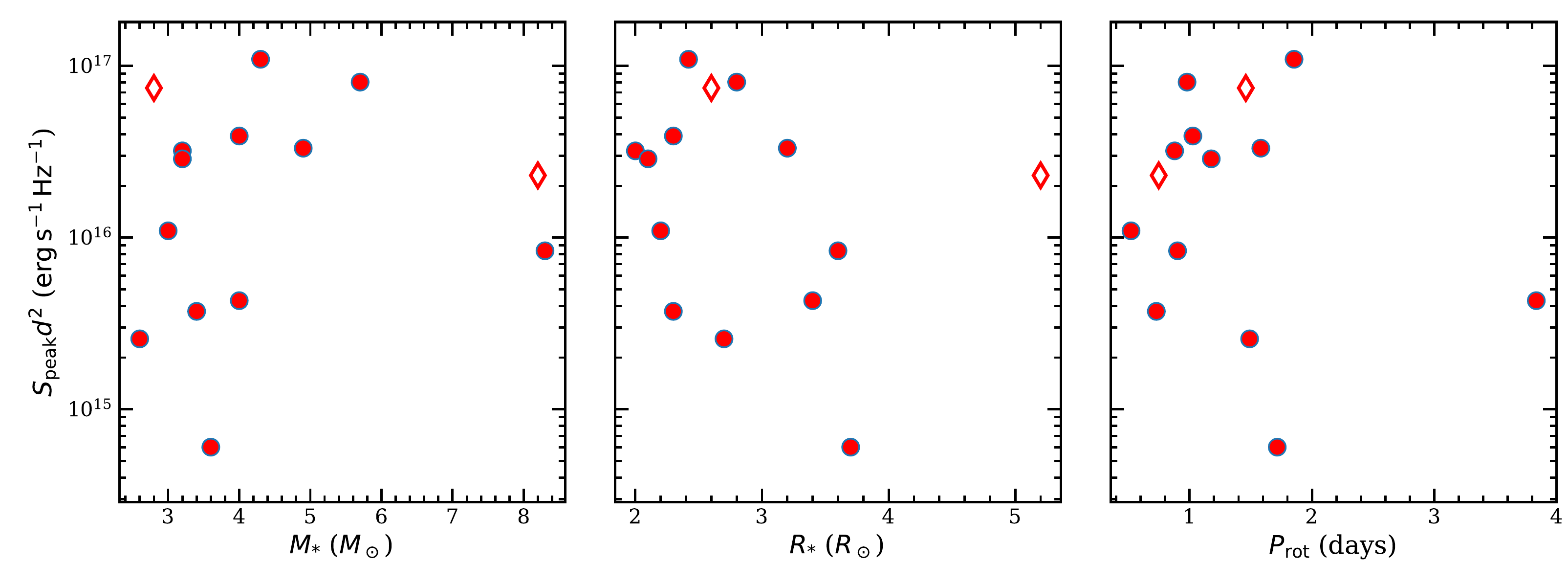}
    \includegraphics[width=0.65\textwidth]{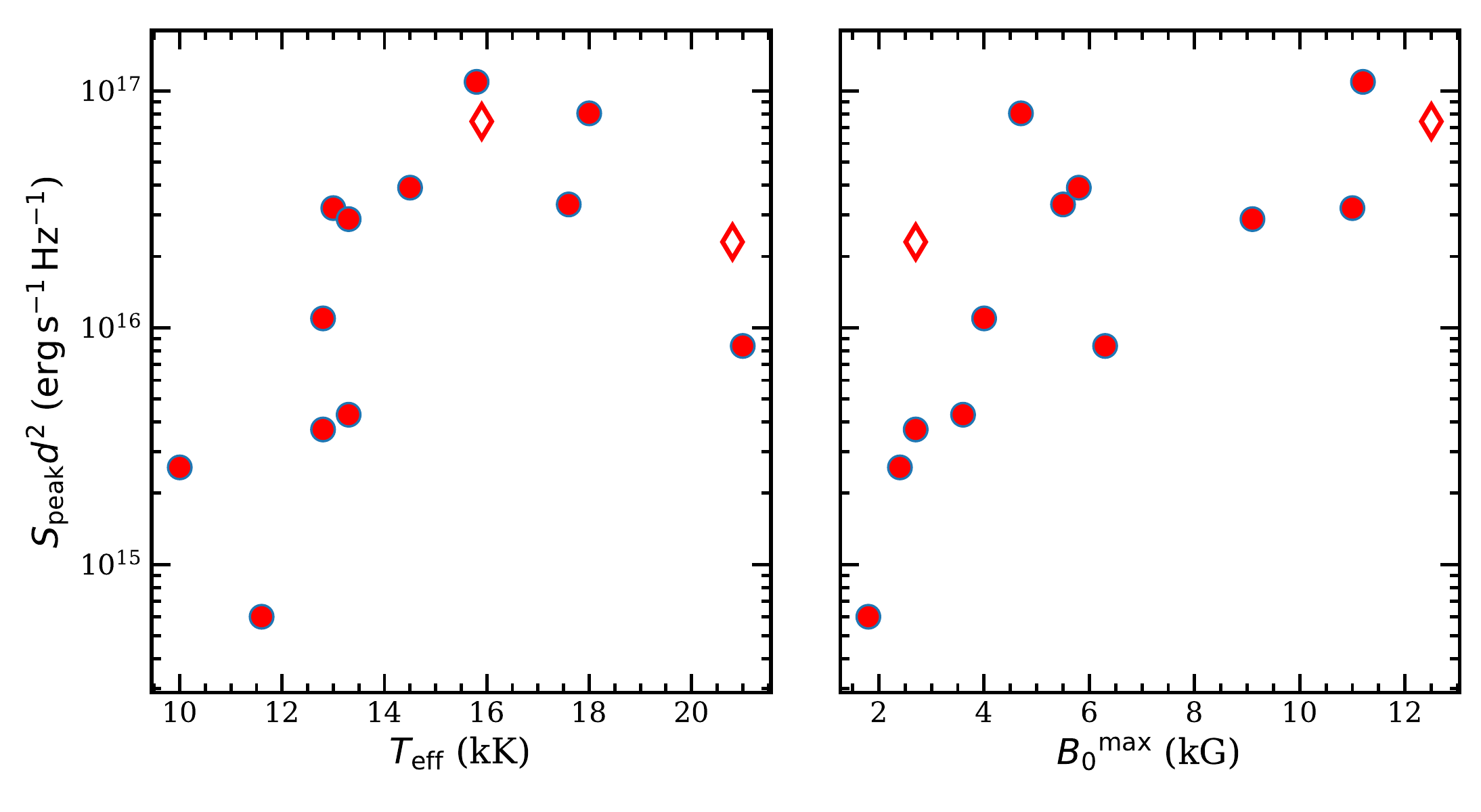}
    \caption{Variation of the quantity $S_\mathrm{peak}d^2$ (proxy for the peak ECME luminosity; \S\ref{subsubsec:ecme_onset}) with stellar mass $M_*$, radius $R_*$, rotation period $P_\mathrm{rot}$, effective temperature $T_\mathrm{eff}$ and the maximum magnitude of the surface magnetic field strength ${B_0}^\mathrm{max}$. The filled circles represent data at frequencies 0.6--0.8 GHz (GMRT data), whereas the two open diamonds, which correspond to the stars HD\,142301 \citep{leto2019} and HD\,147933 \citep{leto2020a}, represent data taken at 1.5 and 2.1 GHz respectively. $S_\mathrm{peak} d^2$ corresponds to the maximum flux density observed from a star at any of the two circular polarizations.}
    \label{fig:lum_vs_MRTBP}
\end{figure*}

As a first step, we compare the luminosity corresponding to the peak flux density of the ECME pulses for all the known MRPs. Note that the peak flux density corresponds to one of the two circular polarizations and not the total intensity. We use the quantity $S_\mathrm{peak}\times d^2$ as a proxy for the peak luminosity, where $S_\mathrm{peak}$ is the `excess' peak flux density (with respect to the basal flux density due to gyrosynchrotron) and $d$ is the distance to the star. 
In the case of HD\,147932 (a.k.a. $\rho\,\mathrm{Oph\,C}$), no estimate for the basal flux density could be obtained, which is due to the fact that the obliquity is likely zero implying that ECME is observable at all rotational phases \citep{leto2020b}. Therefore, HD\,147932 is not included in our analysis.
Barring the stars HD\,142301 (a.k.a 3\,Sco) and HD\,147933 (a.k.a. $\rho\,\mathrm{Oph\,A}$), the peak flux densities correspond to the ECME pulses observed over 0.6--0.8 GHz (i.e. the band 4 of the uGMRT and the 610 MHz of the legacy GMRT). \citet{das2021} recently reported sub-GHz observations for the MRP CU\,Vir. On one of the days of observation, they witnessed a `giant pulse' in band 4 of the uGMRT, which was an order of magnitude brighter than the typical pulses observed from this star. This phenomenon is very likely a transient event and hence, we have not used this pulse in the analysis presented in this paper. However, the qualitative picture does not change even if we include this giant pulse. For HD\,142301, we use its peak flux density at 1.5 GHz reported by \citet{leto2019}. For HD\,147933, we use the peak flux densities at 2.1 GHz reported by \citet{leto2020a}.

We examine the variation of the quantity $S_\mathrm{peak} d^2$ with stellar mass $M_*$, radius $R_*$, $T_\mathrm{eff}$, rotation period $P_\mathrm{rot}$ and the maximum magnitude of the surface magnetic field strength ${B_0}^\mathrm{max}$ \footnote{This is not necessarily the same as the dipole strength; e.g. for the star CU\,Vir, the maximum magnetic field strength is 1--2 kG at the north pole and 4 kG at the south pole \citep{kochukhov2014}. In such a case, we take ${B_0}^\mathrm{max}=4$ kG.} (Figure \ref{fig:lum_vs_MRTBP}). Note that $M_*$ is correlated to $R_*$ and $T_\mathrm{eff}$. Nevertheless, we find the tightest correlation of $S_\mathrm{peak} d^2$ with $T_\mathrm{eff}$ and weakest (or no) correlation with $R_\mathrm{*}$ among the three quantities (Figure \ref{fig:lum_vs_MRTBP}). Similarly, we find that the peak luminosity and ${B_0}^\mathrm{max}$ are correlated. With $P_\mathrm{rot}$, we do not find any correlation of the peak ECME luminosity. This however cannot be trusted since MRPs discovered so far essentially span only a very narrow range of stellar rotation periods.


From Figure \ref{fig:lum_vs_MRTBP}, we find that the relation between the quantity $\log(S_\mathrm{peak} d^2)$ and $T_\mathrm{eff}$ is nearly a parabola with a vertex ($T_0$) around $15-18\,\mathrm{kK}$. To locate $T_0$ more precisely, we calculated the Spearman's rank correlation co-efficient\footnote{Spearman's rank correlation can assess monotonicity of a relation, and is thus more general than the Pearson correlation coefficient which can assess only linear relation. For a perfect correlation or ant-correlation, the correlation coefficient (denoted by $\rho$) is $\pm 1$ with a p-value of zero.} $\rho$ \citep{spearmanr_ref} between the quantities $S_\mathrm{peak} d^2$ and ${(T_\mathrm{eff}-T_0)}^2$, varying $T_0$ between 15 and 18 kK (with a step-size of 0.5 kK). This procedure yields $T_0=16.5\,\mathrm{kK}$. The corresponding value of $\rho$ is $-0.88$ with a p-value of 0.0001. A similar exercise with stellar mass yields $\rho=-0.60$ with a p-value of 0.04. Thus, our preliminary analysis with the sample of 12 MRPs (excluding HD\,142301, HD\,147932 and HD\,147933) suggests that the peak ECME luminosity is a maximum among magnetic hot stars with $T_\mathrm{eff}\approx 16.5$ kK. 
In the case of the correlation with magnetic field strength, the Spearman's rank correlation coefficient 
is $+0.76$ with a p-value of 0.004.

\begin{figure}
    \centering
    \includegraphics[width=0.45\textwidth]{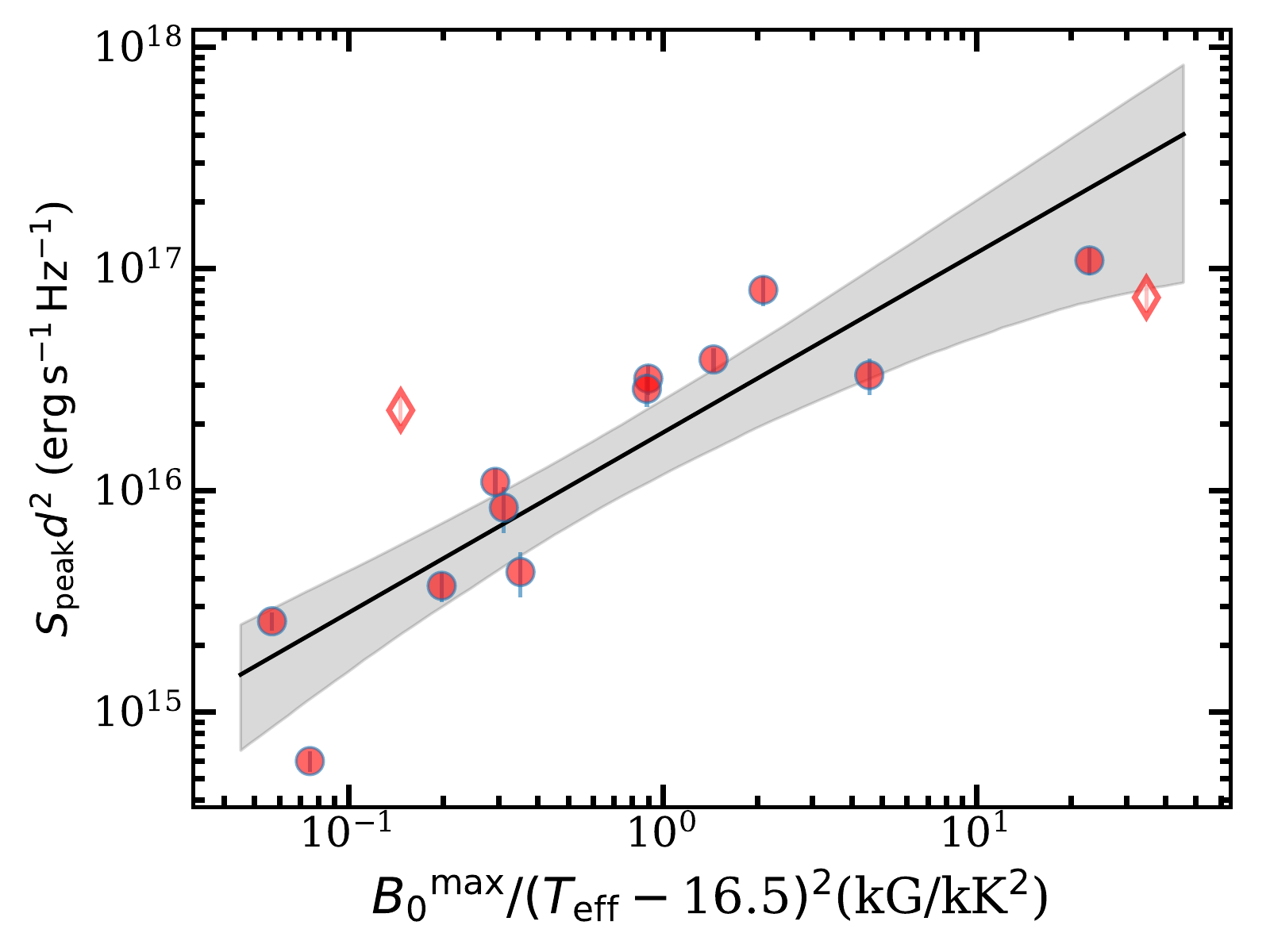}
    \caption{The derived relation between peak ECME luminosity (represented by $S_\mathrm{peak}\times d^2$), maximum magnitude of the surface magnetic field ${B_0}^\mathrm{max}$ and the surface temperature $T_\mathrm{eff}$. The filled circles represent stars for which data at band 4 (0.6--0.8 MHz) is available. The two unfilled diamonds are for $\rho\,\mathrm{Oph A}$ (data corresponds to 2.1 GHz) and HD\,142301 (data corresponds to 1.5 GHz). The solid line corresponds to the relation $S_\mathrm{peak}d^2\propto X^{0.8\pm0.1}$, where $X={B_0}^\mathrm{max}/{(T_\mathrm{eff}-16.5)}^2$. The surrounding shaded regions around the solid line shows the $3\sigma$ uncertainty. The outlier around $X=0.14$ corresponds to HD\,147933.  See \S\ref{subsubsec:ecme_onset} for details. }
    \label{fig:derived_relation_peak_lum_B_T}
\end{figure}

Based on the above results, 
we construct the quantity $X={B_0}^\mathrm{max}/{(T_\mathrm{eff}-16.5)}^2$ and plot the proxy for the peak ECME luminosity against this quantity. Figure \ref{fig:derived_relation_peak_lum_B_T} shows the result. As expected, we find a much tighter correlation between the peak luminosity and $X$ with $[\rho,\mathrm{p}]=[0.94,4\times10^{-6}]$ without including $\rho\,\mathrm{Oph A}$ and HD\,142301. Including the latter two stars slightly deteriorates the correlation giving $[\rho,\mathrm{p}]=[0.90,10^{-5}]$.
By fitting a power law to this relation (again excluding HD\,142301 and HD\,147933), we obtain that the peak ECME luminosity $\propto X^{0.8\pm0.1}$. 
With more discoveries of MRPs, it will be possible to check whether such a relation indeed holds true.

We now consider possible explanations for the observed correlation. At first sight, the correlation between the peak ECME luminosity and the maximum surface magnetic field might look obvious given that the magnetic field is the primary ingredient for triggering ECME. However, other than determining the frequency of emission, the absolute value of the dipole strength does not play a role in the ECME growth rate \citep[e.g.][]{lee2013}. Keeping the observation frequency constant, when we compare ECME luminosity in stars with different dipole strengths, we effectively compare luminosity for ECME produced at different heights from the stellar surface. Thus the correlation with magnetic field strength directly translates to the statement that peak ECME luminosity increases as we go farther from the stellar surface. 
If this statement holds, it should also be reflected in the ECME spectrum of individual stars (in the form of a negative spectral index). However for the few MRPs for which ECME spectra have been reported, such behavior has only been observed close to the upper cut-off frequencies \citep[much higher than the frequency range of band 4,][]{das2020b,das2021}. This rules out the interpretation that the correlation with surface magnetic field strength signifies a dependence on the height of the emission sites from the stellar surface.


There is however an alternate way to view this correlation, which involves the process that energizes the electrons. 
\citet{leto2021} demonstrated that gyrosynchrotron luminosity scales strongly with the unsigned magnetic flux ($B_\mathrm{d}R_*^2$) and rotation. The relation with the latter led them to speculate that centrifugal breakout (CBO) events, previously shown by \citet{shultz2020} and \citet{owocki2020} to regulate H$\mathrm{\alpha}$ emission, may be the source of electron acceleration, a scenario that has been worked out in detail by Shultz et al. (in prep.)
In that case, the stronger the surface magnetic field, the larger the Alfv\'en radius $R_\mathrm{A}$, and the higher the rotational energy of the co-rotating plasma near $R_\mathrm{A}$. Thus, the energy involved in CBO events in a star with a stronger surface magnetic field is very likely higher than that in a star with a weaker magnetic field (Shultz et al. in prep.). With a larger energy reservoir, more electrons can be energized, which might be the root of the origin of the observed empirical relation between peak ECME luminosity and the surface magnetic field strength.



Unlike ${B_0}^\mathrm{max}$, the relation between the ECME luminosity and temperature is non-monotonic. One probable scenario is that for very low $T_\mathrm{eff}$, the stellar wind is weak so that there are not enough particles to emit ECME. For very high $T_\mathrm{eff}$, the wind will be stronger, the magnetosphere will be denser and associated absorption of the radio emission is likely to increase. Besides, with increasing plasma density, it becomes increasingly difficult to maintain the necessary condition of $\nu_\mathrm{p}<\nu_\mathrm{B}$, where $\nu_\mathrm{p}$ and $\nu_\mathrm{B}$ are respectively the plasma and electron gyrofrequencies.

Our finding of the dependence of the peak ECME luminosity on temperature is in stark contrast to the scaling relation obtained for gyrosynchrotron emission by \citet{leto2021}, where the only physical parameters involved are the magnetic flux and rotation period.
This is consistent with the idea proposed by \citet{leto2021} that the incoherent and coherent parts of the radio emission are produced by different populations of electrons and at different sites of the stellar magnetosphere. A rigorous conclusion, however, can only be drawn following a similar analysis with a larger sample of MRPs.

In the past, there has been only one magnetic hot star where ECME was reported to be absent: HD\,37479 \citep[a.k.a. $\sigma$\,Ori E;][]{leto2012}. The lowest frequency observed by \citet{leto2012} was 1.4 GHz. The absence of ECME was attributed to the presence of higher order moments in the stellar magnetic field. It is however yet to be examined whether the star produces ECME at sub-GHz frequencies. For this star, using the available measurements for the stellar ${B_0}^\mathrm{max}$ and $T_\mathrm{eff}$ \citep{oksala2012, shultz2019a,shultz2019b,oksala2015}, we obtain $X=0.24$. From Figure \ref{fig:derived_relation_peak_lum_B_T} and using the known distance to the star \citep{gaia2018}, we find that the peak ECME flux density (if it indeed produces ECME) will be only a few mJy or less. This is consistent with the fact that no ECME pulse has been observed from the star at frequencies $\geq 1.4$ GHz since the basal flux density of the star at 1.4 GHz itself varies between 2--3 mJy \citep{leto2012}. Nevertheless, it will be highly important to observe the star at lower radio frequencies, especially to check the validity of our empirical relation.

Another star that has been observed for nearly a full rotation cycle at higher radio frequencies, but yet to be observed in detail at lower frequencies, is HD\,182180. Radio observations covering nearly the full stellar rotation cycle were reported by \citet{leto2017} at 6--44 GHz. The lowest frequency of their observation (6 GHz) is higher than the typical cut-off frequencies observed for MRPs \citep[e.g.][]{das2020b}. Thus it is not a surprise that ECME was not observed from this star at those frequencies. For this star also, we attempt to examine whether it is likely to produce detectable ECME at lower radio frequencies. We use the measurements reported in \citet{oksala2010,rivinius2010,rivinius2013} to find $X=0.9$. The predicted peak ECME flux density (in excess of the basal gyrosynchrotron flux density) of the star comes out to be $\lesssim 5$ mJy \citep[after we use the distance from][]{gaia2018}. Since at sub GHz frequencies, one does not see detectable modulation of gyrosynchrotron emission, it is a promising candidate to observe at low radio frequencies.

Finally, we consider the star HD\,61556 (a.k.a. HR\,2949) that has been recently detected in the circular polarization survey carried out with the Australian Square Kilometre Array Pathfinder (ASKAP) telescope at 887.5 MHz \citep{pritchard2021}. The reported circular polarization is $76\pm16\%$ \citep{pritchard2021} which makes it a highly likely MRP candidate. For this star, we find $X=0.7$ \citep[values of the stellar parameters are taken from][]{shultz2015,shultz2019a,shultz2019b}. This implies that the peak ECME flux density from this star will be $\sim 10$ mJy  \citep[after we use the distance from][]{gaia2018} which independently predicts that this star is a MRP.


A limitation of this analysis is that the peak flux densities of ECME pulses are known to be variable \citep[e.g.][]{trigilio2011,das2021}. This makes the quantitative prediction of the peak ECME flux density for a given star from our empirical relation somewhat unreliable. However the usefulness of a relation like the one depicted in Figure \ref{fig:derived_relation_peak_lum_B_T} lies in its ability to predict whether a magnetic hot star is likely to produce coherent radio emission or not.

To summarize, our analysis, based on the data for 14 of the 15 MRPs, suggest that the primary physical quantities that determine whether ECME from a star will be detectable or not are the maximum surface magnetic field strength and the surface temperature. While the efficiency of the phenomenon appears to increase monotonically with increasing magnetic field strength, it reaches a maximum around a surface temperature of 16--17 kK. 
To examine the robustness of these inferences (and also to be able to predict whether a hot magnetic star will produce detectable ECME or not), it will be important to continue searching for more of these objects.

\subsubsection{Influence of the obliquity on the ECME pulse profiles}\label{subsubsec:obliquity_influence}
One interesting suggestion that has come out of this work is the influence of the stellar obliquity $\beta$, i.e. the angle between the magnetic dipole and rotation axes, on the ECME pulse-profile. Five stars in our sample have $\beta$ close to $90^\circ$: HD\,12447, HD\,19832, HD\,79158, HD\,145501C and HD\,176582 (Table \ref{tab:targets_properties}). The latter has however only partially covered ECME pulse-profile (Figure \ref{fig:hd176582}) and hence will not be included in this discussion.
Previously, only one MRP, HD\,142990, that has an obliquity close to $90^\circ$ \citep{shultz2019b}, was known \citep{lenc2018,das2019a}. The four stars: HD\,142990, HD\,12447, HD\,19832 and HD\,145501C, exhibit highly peculiar pulse-profiles, characterized by clearly separated sub-pulses (at the same polarization) instead of a single pulse (e.g. compare Figures \ref{fig:hd19832_lightcurves_with_bz} and \ref{fig:hd45583_lightcurves_bz}). HD\,79158 is apparently an exception from this point as we did not observe such feature in its pulses (Figure \ref{fig:hd79158_lighcturves_with_bz}). However, a confirmed inference can only be drawn after we observe the star over the broader rotational phase window. 
The peculiar pulse-profiles in stars with $\beta\approx90^\circ$ is consistent with the simulation results of \citet{das2020a} where they showed that large obliquity might lead to highly non-intuitive ECME pulse-characteristics, purely due to propagation effects in the magnetosphere. Thus, in the future, it will be important to conduct a study including MRPs with such large obliquities (e.g. $>85^\circ$) to understand how this aspect of the stellar magnetosphere influences the ECME characteristics.


Physically one can understand this effect by considering the fact that the plasma distribution in the stellar magnetosphere is a strong function of the obliquity \citep[e.g.][]{townsend2005}. For the case where the rotation and dipole axes are aligned, the distribution (theoretically) is symmetric about the magnetic/rotation axes. It is characterized by the presence of a dense disk in the magnetic/rotational equatorial plane and away from that, the density falls sharply \citep{townsend2005}. However, when the two axes are not aligned, the distribution loses the symmetry. The disk-like overdensity no longer remains at the magnetic equator \citep[e.g. see Figure 9 of ][]{das2020a}. The case when the obliquity is $90^\circ$ is an extreme situation and it is still not clear how the plasma will organize itself in such a case. But according to the semi-analytical `Rigidly Rotating Magnetosphere' (RRM) model of \citet{townsend2005}, in such a case, the disk gets warped with increasing $\beta$
and becomes two cones when $\beta=90^\circ$ \citep[see Figure 3 of][]{townsend2005}.
The double ECME pulses observed from stars with obliquities $\approx90^\circ$ might be a signature of such a density distribution in the stellar magnetosphere.

It is however to be noted that whether or not the ray path corresponding to the observed emission will pass through regions with high density gradients, is likely to be dependent on other stellar parameters as well, like the inclination angle $i$ and the rotation period $P_\mathrm{rot}$. Together with the stellar mass and radius, the latter defines the Kepler radius $R_\mathrm{K}$, which is the distance at which the centrifugal force due to co-rotation balances gravity. According to the RRM model of \citet{townsend2005}, plasma cannot accumulate at distances smaller than $\approx 0.87 R_\mathrm{K}$. Thus for two stars with similar physical parameters except for $P_\mathrm{rot}$, the slower rotator will have a larger $R_\mathrm{K}$, and hence a smaller extent of the plasma disk. For a more rapid rotator, the probability that the observed radiation has to pass through the overdense region will be higher. To summarize, a large misalignment between rotation and dipole axes will not necessarily lead to a peculiar pulse profile; however a peculiar pulse profile is very likely an indicator of large obliquity. Once again, a larger sample of MRPs will be very helpful to disentangle the effects of inclination angle, obliquity and rotation period.

\subsection{Lack of high circular polarization}\label{subsec:lack_of_V}
Our results shows that 100\% circular polarization is not a necessary criterion to identify MRP candidates. In the ideal case of a star with an axi-symmetric magnetic field, the net circular polarization in the observed pulses will be zero when the radiation coming from opposite magnetic hemisphere suffers no deviation at all on its way towards the observer \citep{leto2016}. Most recently, \citet{das2020a} showed via simulation that the radiation may experience significant deviation in the stellar magnetosphere, and yet the net circular polarization can be much smaller (see the bottom panel of their Figure 10). This is consistent with the fact that for the star HD\,12447, the observed circular polarization is quite small, and also the pulse-profiles are peculiar; the latter being indicative of the experience of significant propagation effects in the stellar magnetosphere \citep{das2020a}. 

\subsection{Effect of complex surface magnetic field}\label{subsec:comple_B_field}
Among the eight MRPs that we report in this paper, only two (HD\,12447 and HD\,19832) exhibit \bz~modulation that can be reasonably well-fitted by assuming a dipolar surface magnetic field. But in these two cases, the error bars in the \bz~values are much larger than those for the other stars so that the `good-fit' obtained using a sinusoidal function might be a `limitation' of that. 
Among the already known MRPs, CU\,Vir, HD\,133880, HD\,142990 and HD\,35298 are known to possess magnetic fields more complex than that of an axi-symmetric dipole \citep{kochukhov2014,kochukhov2017,shultz2018}. Thus we do not have any evidence of higher order magnetic moments suppressing ECME. Furthermore, offsets of the rotational phases of arrival of ECME pulses from their predicted values (close to the magnetic nulls) have been attributed to complex surface magnetic fields \citep[e.g.][]{leto2019,das2019a}. We, on the other hand, observed ECME pulses right at the magnetic nulls for several of the eight stars with complex surface magnetic fields (e.g. HD\,45583), but away from the magnetic null for one (HD\,12447) of the two stars exhibiting apparently sinusoidal \bz~variation. Thus, it appears that the offset in the rotational phases of arrival of ECME pulses is not exclusively dependent on the magnetic field topology, but on other stellar physical parameters. For example, in case of HD\,12447, we propose that the star's large obliquity is behind its peculiar pulse characteristics. Sometimes, the offsets could be artificial, e.g. due to the use of an insufficiently precise ephemeris.

One caveat here is that with increasing distance, the higher order magnetic moments decay faster than the dipolar moment, so that below a certain radio frequency, the magnetic field will effectively be dipolar. Since we are comparing stars with different ${B_0}^{\mathrm{max}}$, the same observation frequency corresponds to different heights from the stellar surface. For example, assuming a dipolar geometry and emission at the fundamental harmonic, an observation frequency of 0.7 GHz corresponds to a height of 2.3 $R_*$ for HD\,45583 ($B_\mathrm{d}=9.1\,\mathrm{kG}$, Table \ref{tab:targets_properties}) and 0.9 $R_*$ for HD\,170000 ($B_\mathrm{d}=1.8\,\mathrm{kG}$, Table \ref{tab:targets_properties}) from the stellar surface. However even if we compare stars with similar ${B_0}^{\mathrm{max}}$ (e.g. HD\,12447 and HD\,170000), the discrepancy, described above, remains.

\subsection{Suitable observation strategy}\label{subsec:suit_obs_str}
One important issue raised in this paper is the lack of a strategy to search for this phenomenon which will be suitable for any magnetic hot star. 
Since it is now well-accepted that the radio pulses due to ECME can be visible at rotational phases that are offset from the magnetic null phases, one cannot rule out the existence of the phenomenon by merely observing a star over a specific rotational phase window. Besides, for the MRP CU\,Vir, the ECME pulses at 2.3 GHz have been found to be intermittent \citep{ravi2010}. That is why the ideal way to find out if a star produces ECME or not is to observe the star over as large a rotational phase window (around the magnetic nulls) as possible (preferably for one full rotation cycle). Even then, it remains difficult to completely rule out the phenomenon due to our lack of understanding regarding the ECME cut-off frequencies. 
Based on the currently available data, it appears that ECME is favoured at frequencies $\lesssim5$ GHz. Unfortunately a similar estimate for the lower limit to the suitable observing frequency is not available as barring HD\,133880, the lower cut-off frequency of ECME from other MRPs has not been observed. In the case of HD\,133880, \citet{das2020b} reported that one of the ECME pulses has a tentative lower cut-off frequency at around 0.4 GHz, though for the other pulses, the lower cut-off frequency is clearly below 0.4 GHz. The lowest frequency of observation of ECME from MRPs is 0.2 GHz \citep[HD\,142990,][]{lenc2018}. To the best of our knowledge, no detailed study (e.g. obtaining the lightcurves) has been performed for MRPs below 0.4 GHz.
To understand the low frequency characteristics of ECME, it will be important to study the known MRPs below their current lowest frequencies of observation.

Though we have some handle on choosing the observing frequency, observing each star for one complete rotation cycle to overcome the issue of offset in the rotational phases of arrival of the pulses is practically impossible. This is especially the case for a survey of slowly rotating stars (which can have periods up to several decades). In such a case, one way to proceed will be to choose any one of the stellar magnetic nulls and observe for as much time as possible around it (this strategy was adopted for HD\,19832, HD\,145501C and HD\,45583). The limitation of this strategy is that it might give a false non-detection for stars that, like CU\,Vir, exhibit intermittent ECME pulses. An alternate way for discovering more MRPs is to use sky surveys. The MRP HD\,142990 was first identified as a candidate when it was detected in an all-sky circular polarization survey carried out with the Murchison Widefield Array \citep{lenc2018}. Recently, \citet{pritchard2021} reported the detection of three magnetic chemically peculiar stars (that were not detected in radio previously) in the circular polarization survey conducted with the ASKAP telescope. All three stars are potential MRPs and require targeted observation. This strategy has the disadvantage of excluding MRPs that do not give rise to high circular polarization. More importantly, such surveys are not scheduled to target a particular rotational phase range of individual stars (e.g. around the magnetic null phases), which is likely to affect the detection rate significantly. Nevertheless, for increasing the sample size of MRPs, circular polarization sky-surveys, followed by targeted observation seems to be a useful supplement to targeted observations of well-characterized magnetic stars.


\section{Conclusion}\label{sec:conclusion}
The primary results and the conclusions drawn from this work are listed below:
\begin{enumerate}
    \item More than doubling the sample of MRPs: We report eight new MRPs: HD\,12447, HD\,37017, HD\,19832 HD\,45583, HD\,79158, HD\,145501C, HD\,170000 and HD\,176582. This makes the total number of known MRPs 15. Out of these 8 stars reported here, 10 are discovered and one is confirmed using the GMRT.
    \item ECME is not a rare phenomenon: We find that at least 32\% of magnetic hot stars with physical properties within the ranges spanned by the sample of MRPs, and with visible magnetic nulls exhibit ECME. 
    \item Onset of ECME: For the first time, we perform a comparative analysis using 14 of the 15 MRPs and present an empirical relation to predict whether a hot magnetic will produce detectable ECME. Our preliminary analysis suggests that the efficiency of the phenomenon is primarily controlled by the stellar magnetic field strength and the surface temperature.
    \item $T_\mathrm{eff}$ corresponding to maximum ECME luminosity: Our analysis suggests that the peak ECME luminosity reaches its maximum in stars with $T_\mathrm{eff}\approx16-17$ kK. 
    \item Influence of magnetospheric plasma distribution on the ECME pulse-profile: We find that all three MRPs (HD\,142990, HD\,19832 and HD\,145501C), whose pulses are fully covered by observation and with obliquity$\approx90^\circ$ exhibit `double-pulse' profiles (e.g. see Figures \ref{fig:hd19832_lightcurves_with_bz} and Figure \ref{fig:hd145501_lightcurves_with_bz}). In addition, no other MRP has been found to exhibit this characteristic. Since obliquity plays an important role in shaping the magnetospheric plasma distribution, this observation demonstrates the importance of propagation effects in the stellar magnetosphere on the ECME pulse-profile.
    \item Effect of complex stellar magnetic field: Based on the current data, we do not find any evidence of any definitive role of higher order magnetic moments in suppressing the onset of ECME. Our results also disfavour the idea of complex stellar magnetic fields causing offsets in the rotational phases of arrival of ECME.
\end{enumerate}

Lastly, we would like to reiterate the need to increase the sample size of MRPs. Our work clearly suggests that it is not a rare phenomenon, and the primary difficulty lies in coming up with a suitable strategy to observe these stars. 

\section*{Acknowledgements}
We acknowledge support of the Department of Atomic Energy, Government of India, under project no. 12-R\&D-TFR-5.02-0700. BD thanks David Bohlender for suggesting the star HD\,79158 to observe in radio bands. PC  acknowledges support from the Department of Science and Technology via 
SwarnaJayanti Fellowship awards (DST/SJF/PSA-01/2014-15). MES acknowledges support from the Annie Jump Cannon Fellowship, supported by the University of Delaware and endowed by the Mount Cuba Astronomical Observatory.
GAW acknowledges Discovery Grant support from the Natural Sciences and Engineering Research Council (NSERC) of Canada. 
We thank the staff of the GMRT and the National Radio Astronomy Observatory (NRAO) that made our observations
possible. The GMRT is run by the National Centre for Radio Astrophysics of the Tata Institute of Fundamental Research. The National Radio Astronomy Observatory is a facility of the National Science Foundation operated under cooperative agreement by Associated Universities, Inc. This research has made use of NASA’s Astrophysics Data System.

\bibliography{MRP_candidates_second_revision}
\appendix

\section{Spectropolarimetric Measurements of HD\,37017}\label{sec:spectro_data_hd37017}
Twelve magnetic measurements obtained with the ESPaDOnS spectropolarimeter were previously published by \cite{shultz2018}. ESPaDOnS (\'Echelle Spectropolarimetric Device for the Observation of Stars) is a high-resolution ($\lambda/\Delta\lambda \sim 65,000$) \'echelle spectropolarimeter mounted at the Canada-France-Hawaii Telescope (CFHT). The instrument covers the wavelength range from 370 nm to 1050 nm over 40 overlapping spectral orders. Each spectropolarimetric sequence consists of 4 differently polarized subexposures, which can be combined to yield a circular polarization (Stokes $V$) spectrum, or a null polarization ($N$) spectrum in which intrinsic polarization from the source is cancelled out, thereby giving a measurement of the noise and a check on normal instrument observation. The characteristics of ESPaDOnS and the Libre-ESpRIT reduction pipeline were described in detail by \cite{wade2016}. 

Between 29/10/2015 and 3/11/2015 a further 42 ESPaDOnS observations of HD\,37017 were acquired by the BinaMIcS \citep[Binarity and Magnetic Interactions in various classes of Stars,][]{alecian2015} Large Program. Each observation consisted of $4 \times 50$ sec sub-exposures. The observations were taken on 4 nights, with between 30 and 68 min between the start and end of each set of observations. As this corresponds to between 0.02 and 0.05 of a rotational cycle, observations acquired on a single night were co-added in order to increase the signal-to-noise (S/N). The observation log is given in Table \ref{tab:hd37017_obslog}. 

\begin{deluxetable}{lrrrrrr}
\tablecaption{Observation log of HD\,37017. $\phi_\mathrm{rot}$ represents the rotational phase calculated using the ephemeris given in Table \ref{tab:targets_properties}, $n$ gives the number of spectropolarimetric sequences comprising each measurement. DF gives the Detection Flag, either a Non-Detection (ND) or Definite Detection (DD). \label{tab:hd37017_obslog}}
\tablehead{
\hline
HJD$-$ & $\phi_\mathrm{rot}$ & $n$ & \bz & DF (V) & \nz & DF (N) \\ 
2457300 & &  & (G) &        & (G) &        \\
}
\startdata
25.04535 & 0.19 & 6 &   $-869 \pm 132$ & DD  & $130 \pm 140$ & ND \\
29.09279 & 0.68 & 6 &   $-69 \pm 126$ & DD  & $15 \pm 130$ & ND \\
30.09825 & 0.80 & 6 &   $-720 \pm 136$ & DD  & $120 \pm 140$ & ND \\
31.13575 & 0.95 & 12 &  $-1186 \pm 107$ & DD & $70 \pm  110$ & ND \\
\enddata
\end{deluxetable}

To increase the S/N sufficiently to detect and measure the magnetic field, Least-Squares Deconvolution \citep[LSD;][]{donati1997} mean line profiles were extracted, using the iLSD package developed by \cite{koch2010}. The line list was the same as the one used by \cite{shultz2018}, originally obtained from Vienna Atomic Lines Database \citep[VALD3;][]{piskunov1995, ryabchikova1997, kupka1999, kupka2000,ryabchikova2015}, and then cleaned to remove contaminating Balmer, He, and telluric features. 
Due to the broad stellar lines, LSD profiles were extracted using 7.2 km/s velocity pixels, i.e. four times the usual pixel size; this provided a per-pixel S/N boost of a factor of 2, at the expense of velocity resolution.
Evaluation of False Alarm Probabilities using the method and criteria described by \cite{donati1992,donati1997} found 4/4 Stokes $V$ profiles to be definite detections (DD), and all 4 $N$ profiles to be NDs as expected (Table \ref{tab:hd37017_obslog}).

Since HD\,37017 is a double-lined spectroscopic binary with an 18-day orbit \citep{bolton1998}, it was also necessary to remove the contribution of the non-magnetic companion from the Stokes $I$ profile. Radial velocities of the components were measured using the parameterized line profile fitting routine described by \cite{grunhut2017}, and the line profiles were then disentangled (using the full dataset, i.e.\ including the ESPaDOnS measurements analyzed earlier) using the same iterative procedure as adopted by \cite{shultz2018}. The longitudinal magnetic field averaged over the stellar disk \bz~\citep{mat1989} was measured from the disentangled line profiles in order to quantify the line-of-sight magnetic field strength. The same measurement was performed using $N$, yielding \nz. \bz~and \nz~measurements are given in Table \ref{tab:hd37017_obslog}.

\end{document}